\documentstyle[11pt,epsfig]{article}
\newcommand{\be}{\begin{equation}}
\newcommand{\ee}{\end{equation}}
\newcommand{\beqn}{\begin{eqnarray}}
\newcommand{\eeqn}{\end{eqnarray}}

\newcommand{\spc}[1]{\mbox{\hspace{#1}}}

\newcommand{\fP}{I\!\!P}

\begin{document}
\begin{flushright} CERN-TH/98-67\\  
DESY-98-034\\
DTP-98-02\\
hep-ph/9803497 
\end{flushright}
\begin{center}
\begin{Large}
{\bf An Analysis of Diffraction in Deep-Inelastic Scattering}\\
\end{Large}
\vspace{0.5cm}
J. Bartels$^a$, J. Ellis$^b$, H. Kowalski$^c$ and M. W\"usthoff$^d$ \\
\vspace{0.5cm}
$^a${ II. Institut f\" ur Theoretische Physik,
Universit\" at Hamburg, Germany.}\\ 
$^b$ Theory Division, CERN, Switzerland\\
$^c$ Deutsches Elektronen Synchrotron DESY, Germany \\
$^d$ University of Durham, Department of Physics,
Durham DH1 3LE, UK

\end{center}
\vspace{2.0cm}
\begin{abstract}
We propose a simple parametrization for the deep-inelastic diffractive 
cross section. It contains the contribution of $q\bar{q}$ production to 
both the longitudinal and the transverse diffractive structure functions, and
of the production of $q\bar{q}g$ final states from transverse photons.
We start from the hard region and perform a suitable extrapolation into the
soft region. We test our model on the 1994 ZEUS and H1 data, and confront
it with the H1 conjecture of a singular gluon distribution.
\end{abstract} 

\section{Introduction}
Diffractive events are characterized, in general, by the presence of
large rapidity gaps in the hadronic final state that are not
exponentially suppressed. These are conventionally ascribed to
Pomeron exchange. Diffractive processes in deep-inelastic scattering
(DIS) are of particular interest, because the hard photon in the
initial state 
gives rise to the hope that, at least in part, the scattering amplitude 
can be calculated in perturbative QCD (pQCD).
With the increasing amount of data on diffractive 
DIS~\cite{ZEUS,H1}, we have reached a level of accuracy 
that provides deeper insight into the
nature of the Pomeron and its coupling to partons. 
Rapidity-gap events make up a sizable fraction of all DIS events, and can
only be due to the exchange of some colourless object in the cross channel.
The simplest realization of the Pomeron in pQCD is provided
by two gluons of opposite color~\cite{LN}. More detailed models
based upon the two-gluon picture have been formulated both for
perturbative gluons~\cite{BW} 
and for nonperturbative (massive) gluons~\cite{DL,NZ,D,GLM}:
for an alternative approach, see~\cite{BH}. In this paper
we take the point of view that perturbative QCD provides a reasonable
starting point, as in the more detailed model described 
in~\cite{BW,Wu}. In the present paper, we develop a 
simple parametrization based on this model, which can easily be compared
with experimental data.

The simplest description of diffractive DIS starts from the process of
$q\bar{q}$ production. This process is conveniently described in terms of
light-cone wave functions~\cite{NZ,Mue,BFGMS} in the proton rest frame. 
The light-cone wave function of the photon contains the information 
about the dissociation of the fast-traveling photon
into partons, long before the interaction with the proton occurs. 
At the beginning of the scattering process, the
photon splits into a quark-antiquark pair. At sufficiently large
photon virtuality $Q^2$, the 
quark-antiquark pair radiates additional gluons
before it reaches the proton at rest.
At the time of the interaction, the partonic system is spread over a
transverse area which is comparable with the size of the hadron. 
One expects therefore
that the exchanged Pomeron should be close to the usual soft hadronic 
Pomeron. However, inside the final state of the partonic system,
we expect that 
there are also ``hard" configurations, for which the exchanged Pomeron 
should behave
quite differently. These are final states for which the
partonic system is confined to small transverse distances. Examples are
longitudinally-polarized vector particles~\cite{BFGMS} and high-$p_T$
jets~\cite{BLW,BW2}. In the inclusive measurement of diffractive final states, 
one sums
over both these small-distance and large-distance configurations. So far 
there is no
theoretical framework which allows one to predict the relative magnitudes
of
the ``soft" and the ``hard" components of the diffractive cross section,
which must be determined by experiment~\footnote{It was pointed out
in~\cite{ER} that the pseudo-rapidity cuts imposed in certain early
analyses selected a ``hard" component in diffractive DIS, which must
also be present, at some level, even in analyses without this 
{\it a priori} selection.}. Since the cross section for
the ``soft" component is expected to rise 
weakly with energy for any fixed mass of the diffractive
system, whereas the ``hard" part should rise faster, the energy dependence
of the diffractive cross section may help to determine the relative 
sizes of the two components.

In attempting to formulate a model that interpolates between these
two components, one finds that perturbative models based upon two-gluon 
exchange, which are valid {\em a priori} only for small-size final
states,
allow a smooth extrapolation into the soft region. 
Thanks to gauge invariance and colour cancellation, one does not
encounter infrared singularities in models for $q {\bar q}$ production
or $q{\bar q} g$ production,
i.e., there is no need for an artificial 
cutoff. Moreover, the wave-function formalism can be extended to 
include multi-gluon exchange, which is useful for going beyond the
lowest-order two-gluon exchange. 

Since the first observation of diffractive DIS at HERA, several
attempts have been made to compare the data with QCD-based models. In 
particular, the concept of the Pomeron structure function and its DGLAP
$Q^2$ evolution has been applied~\cite{K,Ko,GS,H1}. In these analyses,
the diffractive cross
section has been assumed to consist only of leading twist, and the contribution
of the longitudinal photon has been disregarded. On the other hand, there is
strong evidence that, for small masses of the diffractive system, the 
longitudinal cross section for $q\bar{q}$ production - although formally
of higher twist - is not small compared to the transverse cross section.
We therefore feel that a more complete analysis of the HERA data should include
the longitudinal cross section, and particular emphasis should be given to
the region of small diffractive masses. As a minimal model, one might consider
just the production of quark-antiquark pairs, which should dominate the
small-mass
region. For somewhat larger masses, the production of an extra gluon
has to be taken into account. In a future step, one would also have to
address the
$Q^2$ evolution of both the transverse and the longitudinal cross
sections.
  
In this paper we propose and test a simple parametrization of the 
diffractive cross section that is motivated by the above considerations.
Stimulated by perturbative QCD, we make an
ansatz that consists of four terms, which model both the transverse and
the longitudinal cross sections for
$q\bar{q}$ and $q\bar{q}g$ production. By treating the overall strength 
of these terms and the exponent of the 
energy dependence as a free parameter, we let the data decide which
fractions of the
diffractive cross section belong to the ``soft" and ``hard" parts. 
After a brief description of the model, we compare this Ansatz with
the ZEUS and H1 1994 data, and finally draw a few conclusions.\\ 

\section{A Parametrization for Diffractive DIS}
\subsection{Theoretical Motivation}
The main variables used for the description of diffractive DIS are
the total hadronic energy $W$ of the $\gamma^*$-proton system and the 
diffractively-produced mass $M$. In the analysis of the diffractive
structure function,
it is convenient to use also the variables   
$\beta$ and $x_{\fP}$. In terms of $W$ and $M$, one has 
$\beta = Q^2/(M^2+Q^2)$ 
and $x_{\fP}=(M^2+Q^2)/(W^2+Q^2)$, where we have neglected the proton
mass and the momentum transfer $t$. To connect these
variables with the Bjorken scaling variable $x_B$, we recall that
$x_B=Q^2/(W^2+Q^2)$, which
immediately leads to $x_B=\beta x_{\fP}$.

Before describing our model in somewhat more detail,
we first make a few general remarks. First, we expect the cross section to
be
dominated by very small $t$ values. After integration over final-state
kinematic variables, the $t$ dependence
and the strength of the coupling of the Pomeron to the proton will be
combined in the overall normalization. Next, the $\beta$ spectrum and the 
$Q^2$-scaling behavior follow from evolution of the final-state partons,
and 
can most easily be derived from the light-cone wave functions of the 
incoming photon. 
This part of our model therefore decouples from the dynamics 
inside the Pomeron. On the other hand, the energy dependence of the Pomeron, 
i.e., the $x_{\fP}$-distribution, can be calculated 
at most only partially 
within perturbative QCD. We therefore leave it as a free parameter. 

We have already indicated that, in the proton rest frame, the 
light-cone wave function formalism 
provides a nice intuitive description of diffractive DIS. 
At leading order, when the photon dissociates into a quark-antiquark pair,
we have a single color dipole with a certain momentum distribution given
by the corresponding wave function. At higher order, more partons are
generated and the initial state can be rather complex. At 
leading-twist level, however, 
the basic structure is again a single color dipole: all
partons (gluons or quarks) but one are located within a small area in
impact-parameter space, i.e., at short relative distances, whereas the
remaining single parton is well
separated. The localized parton subsystem carries color
conjugate to that of the single parton, so that one has
again a color-dipole configuration. The short-distance evolution within
the
parton subsystem factorizes, so one only needs to introduce a wave
function for the momentum distribution of the single parton.
With this simplification, we end up with two basic structures: 
a quark-antiquark dipole and a gluon-gluon dipole, 
the latter appearing only at higher order.

\begin{figure}
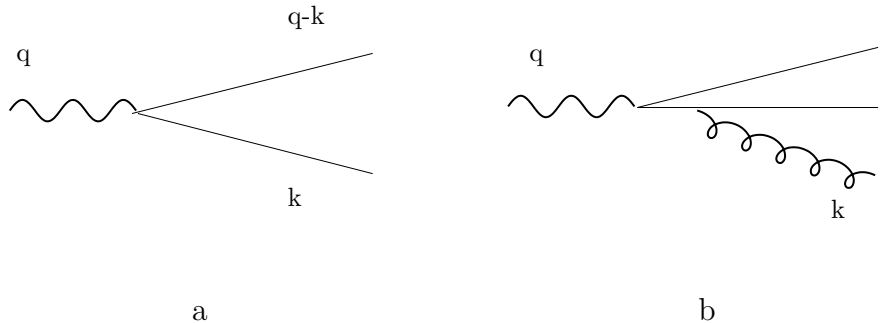

\begin{center}
\input paper55.fig1.pstex_t
\caption{Wave function of the photon, including (a) a $q {\bar q}$
component, and (b) a $q {\bar q} g$ component.}
\end{center}
\end{figure}

In the case of the elementary quark-antiquark final state, the wave
function depends on the helicities of the photon and 
of the (anti)quark. 
We define left- and right-handed transverse photons by projecting
on the polarization vectors (1,i) and (1,-i) ($\gamma=\pm 1$),
respectively, and 
the longitudinal polarization vector is proportional to
the proton momentum $p$ ($\gamma=0$). For massless quarks, the spin is 
orientated along the direction of motion or opposite to it
($h=\pm 1$). As variables for the wave function, we use the 
Sudakov parameters $k=\alpha q'+ \beta_k p +k_t$ with $q'=q+x_Bp$.  
In the proton rest frame with a fast-moving photon, the parameter $\alpha$ 
is of the order unity and denotes the momentum fraction of the photon momentum
carried by the quark (Figs. 1a, 2), whereas $\beta_k$ is small and may be 
neglected. 
Using a complex notation for $k_t$: $k=k_x+ik_y$, $k^*=k_x-ik_y$, one
finds 
for the transverse photon (see also~\cite{Mue}):
\beqn\label{e2.1}
\Psi^{\gamma}_h(\alpha,k_t)
&=&\left\{ \begin{array}{ll}
\begin{displaystyle}
-\;\frac{\sqrt{2}\;(1- \alpha)\;k_t}{|k_t|^2+\alpha (1-\alpha)Q^2}
\end{displaystyle}&\;\mbox{for $\gamma=+1$ and $h=+1$}
\\ & \\ \begin{displaystyle}
\frac{\sqrt{2}\; \alpha\;k_t}{|k_t|^2+\alpha (1-\alpha)Q^2}
\end{displaystyle}&\;\mbox{for $\gamma=+1$ and $h=-1$}
\\ & \\ \begin{displaystyle}
\frac{\sqrt{2}\; \alpha\;k_t^*}{|k_t|^2+\alpha (1-\alpha)Q^2}
\end{displaystyle}&\;\mbox{for $\gamma=-1$ and $h=+1$}
\\ & \\ \begin{displaystyle}
-\;\frac{\sqrt{2}\;(1- \alpha)\;k_t^*}{|k_t|^2+\alpha (1-\alpha)Q^2}
\end{displaystyle}&\;\mbox{for $\gamma=-1$ and $h=-1$}
\end{array} \right. 
\eeqn
Similarly, for the longitudinal photon one finds:
\beqn \label{e2.2}
\Psi^{\gamma}_h(\alpha,k_t)&=& \spc{0.5cm} 2\;\frac{\alpha (1-\alpha) \;Q}
{|k_t|^2+\alpha (1-\alpha)Q^2} \spc{0.48cm} \mbox{for $\gamma=\ 0$ and 
$h=\pm 1$}
\eeqn
Here $Q^2, \gamma$ and $h$ denote the virtuality of the photon and
the 
helicities of the photon and the quarks, respectively, and the wave
function includes
the propagator of the off-shell quark carrying the momentum $k$. 
In the following, we shall make 
use of the small-$k_t$ behavior of (1) and (2). 

To produce a gluon dipole (Fig.1b), we have again to start from the 
approximation described previously,
because a direct coupling of photons to gluons is lacking. 
In the leading-log($Q^2$) approximation, transverse
momenta are strongly ordered, which translates into the inverse ordering
of distances in impact-parameter space. The quark-antiquark pair with a
large
transverse momentum is localized in impact-parameter space, and forms an
effective ``gluon" state conjugate in color to the emitted gluon, which
has a
smaller transverse momentum and is separated by a large distance from
the quark-antiquark pair. In this approach only the transverse
photon polarization is of importance, and it determines the hard part of
the
process, i.e., the effective gluon dipole is independent of whether the
photon
is right- or left-handed. The wave function of the gluon dipole has the
following tensor structure (the indices $\mu,\nu=1,2$, since only the
transverse components are involved):
\be\label{e2.3}
\Psi^{\mu\nu}(\alpha,k_t)\;=\;\frac{1}{\sqrt{\alpha(1-\alpha)Q^2}}\;
\frac{\;k_t^2\;g_t^{\mu\nu}\;-\;2\;k_t^{\mu} k_t^{\nu}}
{\;k_t^2+\alpha(1-\alpha)Q^2}\;\;.
\ee
where $\alpha$ and $k_t$ refer to the diffractively-produced gluon.
We have written explicitly the $(1-\alpha)$ term, even though the
approximation applied here requires $\alpha$ to be much
less than 1, so that $(1-\alpha)\sim 1$. We have introduced this term 
in order to make manifest the
analogy to the previous expression for the quark dipole. Our main interest is,
again, the behavior of (3) near $k_t=0$.
 
\begin{figure}
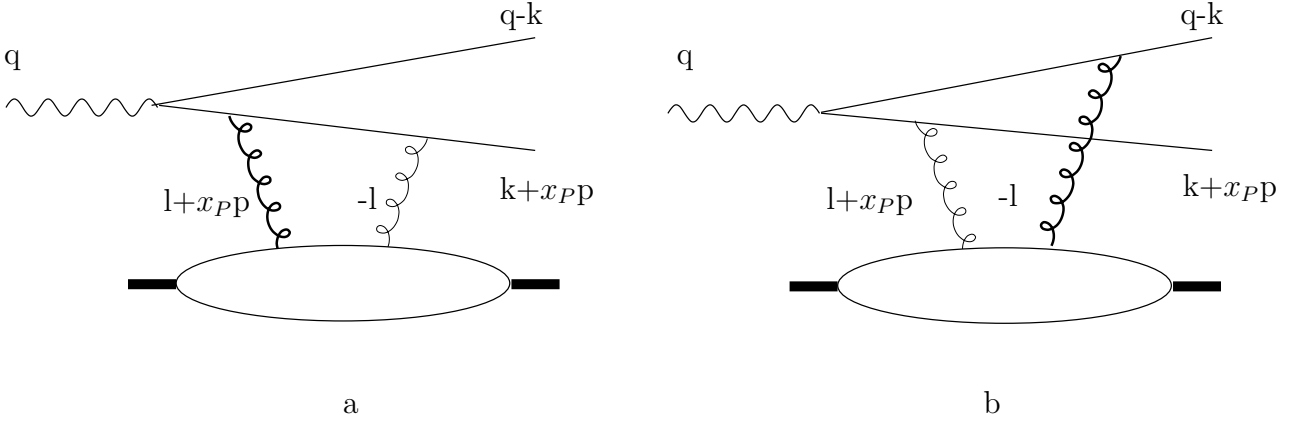

\begin{center}
\input paper55.fig2.pstex_t
\caption{Interaction with the proton, modelled in the two-gluon exchange
approximation.}
\end{center}
\end{figure}
So far we have discussed the photon wave functions which describe the
$q\bar{q}$ or $q\bar{q}g$ state after the splitting of the photon 
into the quark-antiquark pair. In order to obtain the diffractive
scattering amplitude,
we have to include the interaction with the proton target. Beginning with
the $q\bar{q}$ final state, this interaction is represented in Fig.~2. 
It is 
essential that the two gluons couple in all possible ways to the two 
quarks. In the
proton rest frame, the lower quark with momentum $k$ emits a gluon which,
after
the interaction with the proton, is reabsorbed by one of the two quarks. 
A more detailed discussion of this two-gluon-exchange model has been given 
elsewhere~\cite{Wu}. Here we only briefly describe a few main features
that we need 
in order to motivate our parametrization. The coupling of two $t$-channel 
gluons with zero net color and transverse momenta $l_t$ and $-l_t$ to
the color 
dipole, i.e., to any wave function of type (1), (2), or (3), 
can be obtained simply by taking differences of the wave function:
\be
D \Psi \; := \; 2 \;\Psi(\alpha,k_t)\;-\;\Psi(\alpha,k_t+l_t)\;-\;
\Psi(\alpha,k_t-l_t)\;\simeq\;
\left\{ \begin{array}{ll} 
- l_t^i l_t^j\;\frac{\partial^2 \Psi(\alpha,k_t)}{\partial k_t^i
\partial k_t^j}&\mbox{for $l_t\rightarrow 0$}\\
2 \;\Psi(\alpha,k_t)&\mbox{for $l_t \rightarrow \infty$}
\end{array}\right.
\ee
and then convoluting with a suitable ansatz for the $l_t^2$ dependence of the 
Pomeron form factor of the proton.

It is an important feature of our two-gluon model 
that our simple wave 
functions (1), (2), (4), together with a suitable ansatz for the Pomeron 
amplitude, 
provides an interpolation between the hard and the soft region. 
In particular, one finds that the transverse 
polarization belongs to leading twist and is dominated by the aligned-jet
configuration, whereas the longitudinal polarization of the photon leads 
to a higher-twist contribution. In order to
see this behavior in the wave-function formalism, we note the relation 
between the diffractive structure functions $F^D$ and our wave 
functions
\beqn
F^D (x_{\fP}, \beta,Q^2) \;\sim \;\beta \;\int dt\;\int  
\frac{k_t^2 \; d^2k_t}{(1-\beta)^2} \; 
\left|\int \frac{d^2l_t}{l_t^2}
\; D\Psi(\alpha, k_t)\;\phi(l_t^2,k_0^2;x_{\fP}) \right|^2
\eeqn
where $D\Psi$ is taken from (4), and $\phi(l_t^2,k_0^2;x_{\fP})$ stands
for the Pomeron 
amplitude. Here $k_0^2$ denotes some hadronic scale which separates the
regions of 
soft and hard QCD: it should not be confused with the QCD 
factorization scale. The variables $\alpha$, $\beta$, 
$k_t^2$, and $M^2$ are related through
\beqn
\alpha(1-\alpha)M^2 = k_t^2,
\eeqn
and $\beta=Q^2/(Q^2 + M^2)$.
A simple choice for the $l_t$ dependence is~\cite{BLW} 
\beqn 
\phi \;\sim \; \frac{1}{k_0^2}\;\left(\frac{k_0^2}{l_t^2}\right)^
{\nu(l_t^2/k_0^2)}
\eeqn
where $\nu(l_t^2/k_0^2) \approx 1$ as $l_t^2\gg k_0^2$, and $\nu(l_t^2/k_0^2)
\to 0$ as 
$l_t^2\to 0$. 

We start in that part of the kinematic region where the 
parton model of Fig. 2 is most reliable, i.e., the region where the
virtuality of the quark with momentum $k$ is large: 
$k_t^2 + \alpha (1-\alpha) Q^2 = k_t^2/(1-\beta) > k_0^2$~\cite{BLW}.
In this final-state configuration, the $q \bar{q}$ pair has a small 
transverse size, and the two-gluon 
Pomeron interacts with the whole system. Consequently, both contributions
of Fig. 2 are important, and the simple picture of a ``Pomeron structure
function``, which would be
suggested if only Fig. 2a were taken into account, does not
apply~\footnote{This point has previously been emphasized in~\cite{ER}.}.
In this region, the Pomeron amplitude $\phi(l_t^2,k_0^2;x_{\fP})$ coincides, 
to a good
approximation, with the unintegrated gluon structure function of the proton:
\beqn
\int^{k_t^2/(1-\beta)} dl_t^2 \phi(l_t^2,k_0^2;x_{\fP}) =
x_{\fP} g(x_{\fP}, k_t^2/(1-\beta))
\eeqn
(for the extrapolation into the region of smaller $k_t$ and for a discussion
of the dependence on $x_{\fP}$, see below).
Going into more detail, let 
us look into the dependence upon $k_t^2$ and $\alpha$ at fixed $Q^2$. In the 
$l_t$ integral, it suffices to note that the dominant region is 
$k_0^2 < l_t^2 < k_t^2+\alpha(1-\alpha)Q^2$: in our example (7), 
$\phi\sim 1/l_t^2$, and the
dominant contribution, in fact comes from this kinematic domain. In this 
region, we 
approximate $D\Psi$ in (4) by the limit $l_t \to 0$, and obtain:
\beqn
\int_{k_0^2}^{k_t^2 + \alpha(1-\alpha)Q^2} 
\frac{d^2l_t}{l_t^2} \;\phi(l_t^2,k_0^2;x_{\fP}) D\Psi \;&\sim& 
\;\frac{\alpha (1-\alpha)|k_t| Q^2}{(k_t^2 + \alpha(1-\alpha)Q^2)^3}
\; x_{\fP} g(x_{\fP},k_t^2 + \alpha(1-\alpha)Q^2 ) \nonumber \\
&=& \frac{\beta(1-\beta)^2}{k_t^2 |k_t|} x_{\fP} g(x_{\fP},k_t^2/(1-\beta)) 
\eeqn
Inserting this into (5) and making use of relation (6), we find that the 
integral over $k_t^2$ (at fixed $\beta$)
is dominated by the lower limit $k_0^2$. In terms of the variable
$\alpha$, this lower limit corresponds to $\alpha \sim
k_0^2/Q^2$ or $1-\alpha \sim k_0^2/Q^2$. The final result
for (5) is constant in $Q^2$, i.e., it is of leading twist.
The end points of the
$\alpha$ integral correspond to the aligned configuration: in the 
center-of-mass system
of the quark-antiquark pair (Fig.3), $\alpha$ is related to the scattering 
angle
$\theta$ through $2\alpha=1-\cos \theta$, and the dominant regions are
$\theta=0,\pi$. In other words, starting in the hard region of large
transverse momenta, we find that the main contribution comes from the lower
end of the $k_t$ integral, i.e. we find ourselves pushed into the soft
region where our perturbative ansatz for the Pomeron becomes invalid.

A similar argument applied to the longitudinal case shows that the 
dominance of small $k_t^2$ (or values of $\alpha$ close to zero or one) 
is less pronounced: instead of (9), we now have
\beqn
\int_{k_0^2}^{k_t^2 + \alpha(1-\alpha)Q^2} 
\frac{d^2l_t}{l_t^2} \;\phi(l_t^2,k_0^2;x_{\fP}) D\Psi \;&\sim& 
\;\frac{\alpha (1-\alpha) \sqrt{Q^2}(\alpha (1-\alpha)Q^2 -k_t^2)}
{(k_t^2 + \alpha(1-\alpha)Q^2)^3}
\; x_{\fP} g(x_{\fP},k_t^2 + \alpha(1-\alpha)Q^2 ) \nonumber \\
&=&
\frac{\beta (1-2\beta)(1-\beta)}{k_t^2 Q} 
x_{\fP} g(x_{\fP},k_t^2/(1-\beta)).
\eeqn
Inserting this into (5), we see that for small $k_t^2$ the integral
diverges only logarithmically. Also, in contrast 
to the transverse case, the result is of order $1/Q^2$ and hence
belongs to nonleading twist. On the other hand, the integration over 
$k_t^2$ now yields an additional logarithm in
$Q^2/(4\beta k_0^2)$, which is absent in the transverse leading-twist 
case, and slightly compensates for the $Q^2$ suppression. 

Next we return to the transverse case and take a closer look at the soft 
region where $k_t^2 + \alpha (1-\alpha) Q^2 < k_0^2$, 
i.e., $k_t^2 < (1-\beta) k_0^2$ and $\alpha< \beta k_0^2/Q^2$ or 
$1-\alpha<\beta k_0^2/Q^2$. Now
the quark with momentum $k$ in Fig. 2a, before it interacts with the
two-gluon Pomeron, is nearly on shell, and one expects the
picture of the Pomeron structure function to become valid, i.e.,
the lower parton in Fig. 2a can be considered more as a ``valence"
constituent of the Pomeron, and the contribution
of Fig. 2b should be less important. This transition
\begin{figure}
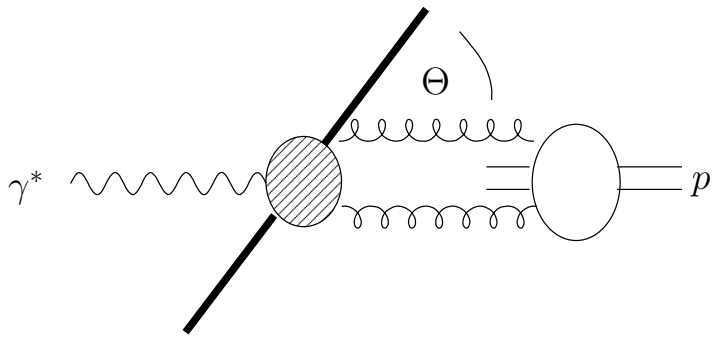

\begin{center}
\input paper55.fig3.pstex_t
\caption{Two-jet production in the $\gamma^*$-$\fP$ center-of-mass system}
\end{center}
\end{figure}
can be made explicit by changing the $l_t^2$ dependence of the
Pomeron amplitude, which now can no longer be identified with the
unintegrated gluon structure function, in such a way that is gives more
weight 
to the region $l_t^2<k_0^2$: the simple example in eq.(6) leads to
$\phi \sim 1/k_0^2=$ constant. The correct behavior is
obtained simply by taking in (4) $l_t^2$ much larger than
$k_t^2$, i.e., $l_t \rightarrow \infty$: 
in this region the first term $2\;\Psi(\alpha,k_t)$ dominates, and we 
end up with 
\beqn
\int \frac{d^2l_t}{l_t^2} \;\phi(l_t^2) D\Psi \;&\sim& \;
\frac{|k_t|}{k_0^2 (k_t^2 + \alpha(1-\alpha)Q^2)}.
\; \ln\left(\frac{k_0^2}{k_t^2 + \alpha(1-\alpha)Q^2} \right) \nonumber \\
&=& \frac{1-\beta}{k_0^2 |k_t|} \ln\left(\frac{(1-\beta)k_0^2}{k_t^2}\right)
\eeqn
which corresponds to the planar diagram Fig. 2a (on the rhs of (11), we have 
disregarded the dependence upon $x_{\fP}$). Returning to (5), the 
integral in $k_t^2$ can easily be performed and leads 
to a finite leading-twist result. By similar arguments the longitudinal case 
is found to be of the order $1/Q^2$. 

In summary, with a suitable Ansatz for the $l_t$ dependence of
the two-gluon 
Pomeron or, even more simply, with a simple prescription for the $l_t$ 
integral, 
it is possible to interpolate between the hard region where the parton model 
applies and the soft region where the aligned jet configuration dominates. 
For the latter case we seem to have arrived at the same conclusions as 
Bjorken~\cite{Bj}, although with a somewhat different line of arguments. 
But, as we shall argue further below, there is a new element that we have
to 
take into account, namely the observation of the strong rise of the
gluon structure function at small x, which gives a weight to the
region of large transverse momenta which is larger than was anticipated 
before the advent of the HERA data in, for example,~\cite{Bj}.

Before we turn to the $x_{\fP}$ dependence of the cross section, let us
mention that the momentum dependence of the $q\bar{q}g$ final state
is quite analogous to that of the $q\bar{q}$ system. We do not show the
analogue of Fig.2: it is again essential that the two gluons couple to the
diffractive system in all possible ways. As before, one has to start
in the region where the transverse momenta of all three partons are large.
When trying to integrate over the transverse momentum of the gluon, one
finds
dominance by the low-momentum region: in this region, all nonplanar
couplings
of the two gluon lines of the Pomeron to the diffractive state become less 
important, and we
are left with the leading-twist `Pomeron structure function' picture,
where the Pomeron 
interacts only with the gluon and not the quarks.  

Next we adress the $x_{\fP}$ dependence of the cross section.
So far we have drawn a rather simple picture of the diffractive final state:
both for the $q\bar{q}$ and the $q\bar{q}g$ final state,
we have argued that in the preferred configuration at least one of the 
final-state partons has a rather soft transverse momentum, and it is 
this parton which couples to the Pomeron. If the virtuality of this
parton is characterized by a typical hadronic scale $\sim \Lambda_{QCD}$,
this seems to 
imply that the energy dependence of the diffractive cross section should
be the same as in hadron-hadron scattering, i.e., the 
diffractive structure function $F_2^D$ grows as $(1/x_{\fP})^{n_{\fP}}$ with
$n_{\fP}=2\alpha_{\fP} (0) -1 \approx 1.12$. However, because of the 
observed rise 
of the gluon structure function at small $x$,  the situation is 
more complicated. Let us return to the above discussion of the 
$q\bar{q}$ final state. The perturbative region is that of large transverse 
momenta of
the final-state partons. For this part of the phase space, we expect the
Pomeron to be described by the perturbative two-gluon model, i.e., the
$x$ dependence of the cross section will be given by the square of the
gluon structure function of the proton~\cite{BLW}: 
\beqn
\frac{d\sigma}{dM^2 dt dk_t^2} \sim [x_{\fP} g(x_{\fP}, k_t^2/(1-\beta)) ]^2
\eeqn
This should lead to a rise $F_2^D \sim
(1/x_{\fP})^{n_{hard}}$ where $n_{hard}= 2\alpha_{hard}-1$ grows with the 
transverse 
momentum $k_t^2$ of the partons, and typically lies above the value $1.4$.
For the kinematic region where the quark
transverse momenta are small and our perturbative two-gluon Pomeron has to be
replaced by some model for the nonperturbative Pomeron, we expect a smaller
exponent $n$: the conventional soft Pomeron would suggest that $n =n_{\fP} 
=2 \alpha_{\fP}(0)-1 \approx 1.12$. Since in the diffractive cross section we
integrate over both the perturbative and nonperturbative parts of the
phase space, there will be competition between the two
regions. At first sight, the large-momentum region seemed to be
rather subdominant. However, the large gluon structure function provides
an
enhancement of this region, and in this way weakens the dominance of the
soft nonperturbative region. As a result,
the effective value of the exponent $n$, $n_{eff}$, is expected to lie 
somewhere between the soft and the hard values, and the effective scale
at which the $k_t$ integral peaks should be somewhat higher than the
soft Pomeron scale. Theoretical studies ~\cite{BLW1} 
indicate that $n_{eff}$ only weakly depends upon $Q^2$, but they do not
allow us to predict the numerical value of $n_{eff}$ or the momentum
scale.  

\subsection{The Parametrization}
After this brief theoretical review, we are ready to 
describe and motivate our parametrization. It will be given in terms of 
the diffractive structure function $F_2^D$, and can be written as the sum of 
several distinct contributions. In our fit we include the following
four pieces:
\beqn
F_2^D=F_{q\bar{q}}^T + F_{q\bar{q}g}^T + \Delta F_{q\bar{q}}^L
+\Delta F_{q\bar{q}}^T.
\eeqn 
Here the first and the 
second term, as indicated by the subscripts and by the superscripts,
denote the production of a quark-antiquark 
pair and the production of a quark-antiquark-gluon system 
with transversely-polarized photons. The third term takes into account 
the production of a quark-antiquark pair from a 
longitudinally-polarized photon, and the prefix $\Delta$ indicates that
this contribution belongs to higher twist (twist four). 
We have also included a  
transverse  higher-twist contribution to $q\bar{q}$ production,
denoted by $\Delta F_{q\bar{q}}^T$

Let us discuss these terms in more detail. To begin with the $Q^2$
dependence of (11), we recapitulate that in the leading-twist transverse 
contribution
to $q\bar{q}$ production there is no $\log(Q^2/Q_0^2)$-enhancement from
the phase-space integral, whereas $q\bar{q}g$ production is of higher
order in $\alpha_s$ and has
an $\alpha_s \ln(Q^2/Q_0^2)$ dependence. The third term, the longitudinal 
cross section of the $q\bar{q}$ final state, belongs to higher twist, and the 
phase-space integral provides a 
$\log(Q^2/Q_0^2)$ enhancement. The reason why this contribution is
essential will be discussed below. The longitudinal contribution of the
$q\bar{q}g$ final state is again of leading twist, but the logarithm
is absent.
It is therefore subleading in comparison with the transverse contribution,
and will be disregarded in our parametrization.
 
To get an estimate of the $\beta$ spectrum, we consider the limits
$\beta \rightarrow 1$ and $\beta \rightarrow 0$. Contact with the other
variables is made through the kinematic relation (6). For the $\beta$ 
dependence of the longitudinal 
cross section we can use (10) and (5): for the 
transverse case the situation is
slightly more complicated, and we have to use both (9) and (11),
in combination with (5). A more
intuitive argument can be derived from the wave functions, and goes as
follows.
The limit $M \rightarrow 0$, which is the same as $\beta \rightarrow 1$,
is related to the small-$k_t$ behavior of the cross section. Let us return to
the wave function (4). For both of the limits $l_t \rightarrow 0$ and $l_t
\rightarrow \infty$, one finds that when $k_t$ approaches zero
$D\Psi$ vanishes with the same power
in $k_t$ as the orginal wave function. (In eq.(\ref{e2.3}),
where the denominator is quadratic in $k_t$, the integration over the
azimuthal angle of $l_t$ leads to the final cancellation of nonvanishing 
contributions.) This means that characteristic features of the light-cone 
wave functions ($\Psi^0, \Psi^{\pm}, \Psi^{\mu\nu}$) remain unchanged after 
scattering. We expect that these results also hold for multi-gluon exchange. 
It is important to note that a single gluon (or photon) exchange, as
opposed to
the color-singlet two-gluon exchange, leads to a rather different
spectrum at $\beta \sim 1$. This is because, instead of the second-order 
derivative in (4), in this case only the first derivative of the wave 
function is needed, which, unlike the second derivative, does not 
vanish when $k_t$ approaches zero. 

Applying these arguments to our wave functions we find, first 
for the the transverse quark-antiquark production cross section, that
the cross section behaves like $(1-\beta)$ ($\Psi^{\pm} \sim Q \vec{k_t}$ from
(\ref{e2.1}) and $M \sim k_t$ from (6)), i.e., it vanishes when $M$
becomes zero. For the second contribution with one gluon in the final state
and with the tensor structure $\Psi^{\mu\nu} \sim k_t^{\mu}k_t^{\nu} -
2|k_t|^2g_t^{\mu\nu}$, the cross section vanishes like $(1-\beta)^2$. The
subsequent integration over the quark-antiquark final state introduces a
further suppression, leading to a $(1-\beta)^3$ behavior. Finally,
for the wave function (\ref{e2.2}) of longitudinally-polarized photons, 
we find that $\Psi^0\sim Q$, i.e., the cross section goes to a constant 
different from zero: this means that near $\beta=1$ the longitudinal cross
section dominates and cannot be neglected. The other limit
$\beta \rightarrow 0$ or $M \rightarrow \infty$ is dictated by the
high-energy behavior of the amplitudes, which is different for quark and
gluon exchange. Spin-1/2 exchange is suppressed relative to spin-1
exchange, which leads to a dominance of gluon production at small $\beta$
over the leading-order quark-antiquark production. We conclude that our
three contributions, transverse
$q\bar{q}$ and $q\bar{q}g$ production and longitudinal $q\bar{q}$ production,
are important in rather
distinct regions in $\beta$, namely medium, small, and large $\beta$,
respectively. The transverse higher-twist contribution is expected to give
a 
small negative correction which is due to phase space limitations at
finite $Q^2$~\cite{BWht}. 

Finally, we comment again on the energy dependence. In contrast to the
$\beta$ spectrum, which can be traced back to rather
general properties of the wave functions, perturbative QCD does not allow
us to control the $x_{\fP}$ dependence of the cross section. In particular,
$n_{eff}$ for the leading-twist transverse cross sections
cannot yet be predicted, and we therefore let the data
decide on the preferred values of this exponent. As discussed at the end of 
the previous subsection, we expect a weak $Q^2$ dependence.   
For the higher-twist longitudinal part, on the other hand, theoretical 
arguments have been given which indicate that
the $x_{\fP}$ dependence is given by the square of the gluon structure
function at the momentum scale $Q^2/4\beta$, i.e., it should grow with
$Q^2$.
In our fit we assume a universal $x_{\fP}$ dependence of all 
higher-twist terms. The exponent is allowed to vary with $Q^2$.

After these remarks, we finally write down our Ansatz for the diffractive
structure function. For the four terms in (11) we put:  
\beqn
F_{q\bar{q}}^T &=& A\;\left(\frac{x_0}{x_{\fP}}\right)^{n_2}\;\beta(1-\beta) 
\nonumber \\
F_{q\bar{q}g}^T &=& B\;\left(\frac{x_0}{x_{\fP}}\right)^{n_2}\;\alpha_s
\;\ln\left(\frac{Q^2}{Q_0^2}+1\right)\;(1-\beta)^{\gamma} \nonumber \\
\Delta F_{q\bar{q}}^L&=& C\;
\left(\frac{x_0}{x_{\fP}}\right)^{n_4}\;\frac{Q_0^2}{Q^2}\;
 \left[\ln \left(\frac{Q^2}{4 Q_0^2 \beta}+1.75 \right)\right]^2\;
\beta^3 (1-2 \beta)^2 \\
\Delta F_{q\bar{q}}^T &=& D\;
                \left(\frac{x_0}{x_{\fP}}\right)^{n_4}\;\frac{Q_0^2}{Q^2}\;
                \ln\left(\frac{Q^2}{4 Q_0^2 \beta}+1.75 \right)
                \beta^3 (1-\beta). \nonumber \\
\eeqn
The exponents are chosen to have the form
\beqn
n_2 &=& n_{2\;0} + n_{2\;1} \ln\left[\ln\left(\frac{Q^2}{Q_0^2}\right)+1\right]
\nonumber \\
n_4 &=& n_{4\;0} + n_{4\;1} \ln \left[\ln\left(\frac{Q^2}{Q_0^2}\right)+1
\right],
\eeqn
and $\alpha_s$ is set to 0.25.

As we have discussed before, the shapes of the 
$\beta$ spectra are restricted by properties of the light-cone wave 
functions, and one finds 
for the parameter $\gamma$ in $F_{q\bar{q}g}^T$ the value 3. 
But, since the H1 analysis
in~\cite{H1} reports a rather hard gluon distribution inside the
Pomeron, we allow the parameter $\gamma$ to deviate from 
the model prediction.  For the
leading-twist (transverse) cross sections the exponent $n_2$
will be left to the fit. We have introduced a simple $Q^2$-dependent
function, although 
a weak dependence is expected theoretically.
The $x_{\fP}$ dependence of the-higher twist contributions
($n_4$), on the other hand, is expected to be
given by the square of the gluon structure function, and in principle we 
could use the measured gluon structure function.  We have inserted
an extra log$(Q^2/(4\beta Q^2_0))$ in order to simulate the effect of 
having a structure function.
For $n_4$ we assume the
same functional form as for $n_2$, but expect to find a stronger rise
with $Q^2$ than for the leading-twist contributions.
The parameter $x_0$ was introduced to
minimize any effect from $n_2$ and $n_4$ on the overall $Q^2$ shape.
The scale parameter $Q_0^2$ is taken to be $1 GeV^2$. 

Finally, we mention that, in principle, one should also allow for
higher-twist corrections to $q\bar{q}g$ production. 
However, even if we assume a 
universal $x_{\fP}$ dependence for all higher-twist pieces, this would 
increase the number of free parameters. We have found that, with the 
presently available amount and accuracy of data, it is not yet possible to 
determine these parameters with sufficient accuracy. Therefore, these
higher-twist pieces will not be included in our fit.

\section{Fits to the Data} 
{\bf ZEUS Data:} We begin with a fit to the ZEUS data
~\cite{ZEUS}, whose results are shown in Table 1
and Fig.4.
\begin{table}[h]
\begin{center}
\begin{tabular}{|c|c|c|c|c|}   \hline
  & mean value & error & lower limit & upper limit \\ \hline
A & 293 & 11 & no & no\\ \hline
B & 166 & 25 & no & no \\ \hline
C & 76 & 15 & no & no \\ \hline
D & -184 & 2 & no & no \\ \hline
$x_0$ & 0.001& 0.00011& 0.0001 & 1 \\ \hline
$\gamma$ & 4.3 & 0.9 & no & no \\ \hline
$n_{20}$ & 1.11 & 0.14& 1 & 2 \\ \hline
$n_{21}$ & 0.12 & 0.12& 0 & 1 \\ \hline
$n_{40}$ & 1 & 0.79 & 1 & 2 \\ \hline
$n_{41}$ & 0.43 & 0.12& 0 & 1 \\ \hline
\hline$\chi^2$/d.o.f.   & \multicolumn{4}{c|}{12/43}\\  \hline
\end{tabular}\vspace{0.5cm}\\
Table 1
\end{center}
\end{table}
\begin{figure}
\begin{center}
\epsfig{file=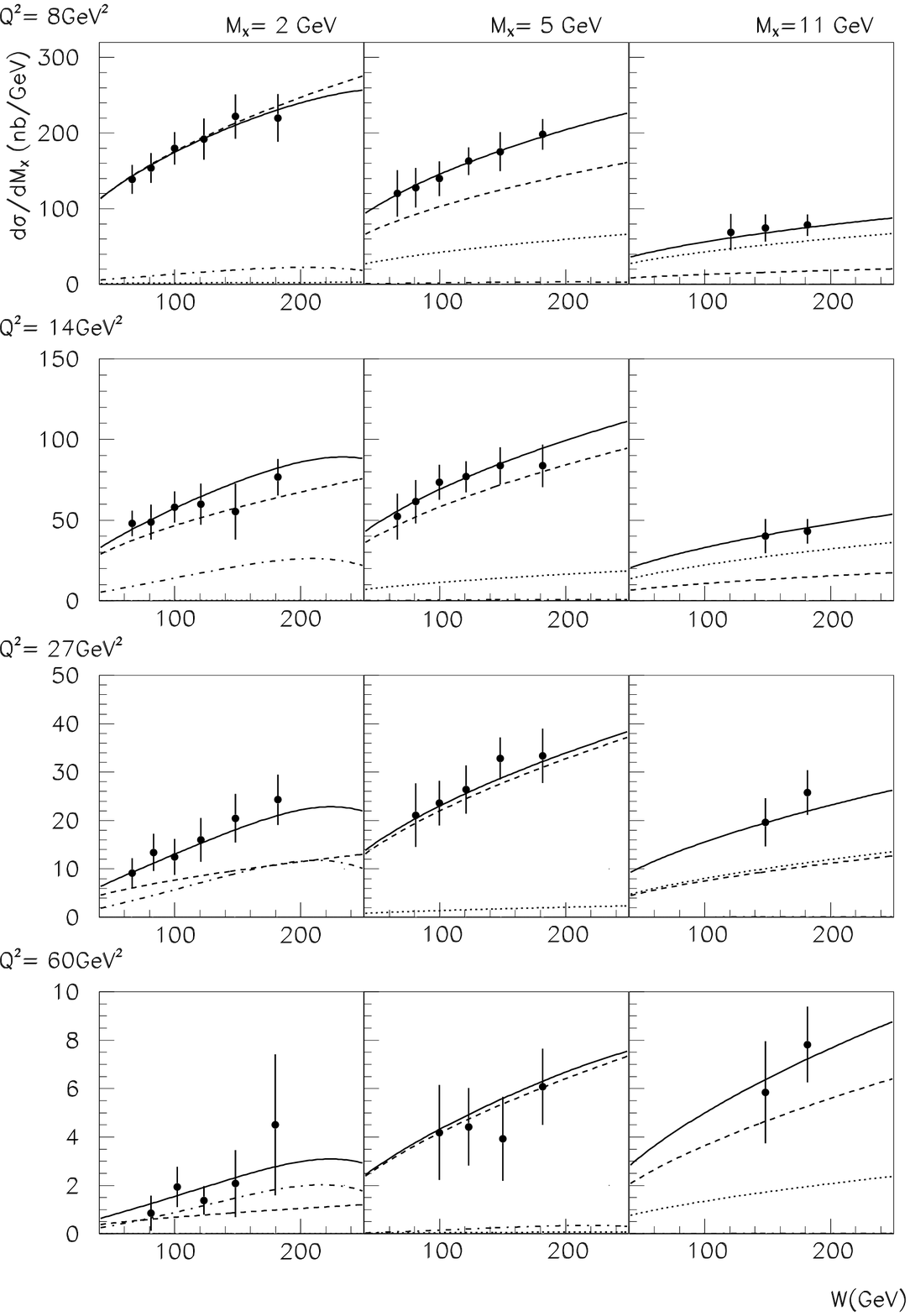,width=14cm}\\
\end{center}
Figure 4: Fit compared to 1994 ZEUS data. Upper solid line: total result, 
dashed line: $F_{q\bar{q}}^T$,
dotted line: $F_{q\bar{q}g}^T$ and dashed-dotted line: $F_{q\bar{q}}^L$.
\end{figure}
Some of the parameters have been restricted
to a physically meaningful range by imposing upper and lower limits. 
We note that the form of the $q\bar{q}g$ wave function in (3) leads
to $\gamma=3$: our fit, cf. the value of $\gamma$ in Table 1 and the
dotted curves in Fig.4, indicates that the 
data prefer a larger value. The behavior of $F_{q\bar{q}g}$
near $\beta=1$ is therefore far from a
`hard gluon distribution' in the Pomeron. Our fit shows that
it is possible to describe the ZEUS data without introducing a hard gluon
inside the Pomeron. 

\begin{center}
\mbox{\epsfig{file=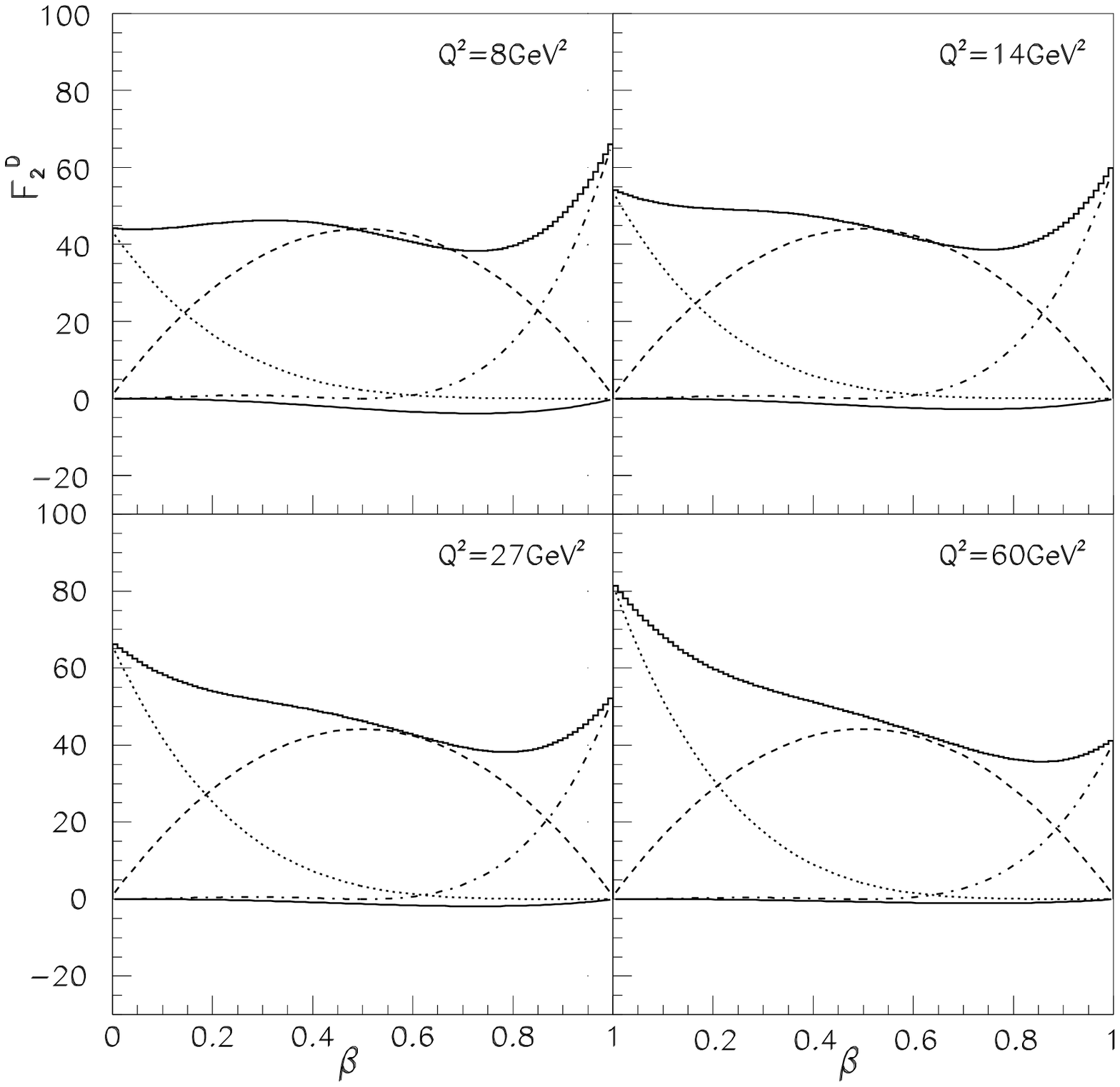,width=10cm}}
\raisebox{5cm}{
\parbox{5cm}{Figure 5: The $\beta$ spectrum at fixed $x_{\fP}=0.001$. 
Upper solid line: total result, 
dashed line: $F_{q\bar{q}}^T$,
dotted line: $F_{q\bar{q}g}^T$, dashed-dotted line: $F_{q\bar{q}}^L$ and
lower solid line: $\Delta F_{q\bar{q}}^T$.}
}\vspace{0.3cm}\\
\end{center}  

Fig.5 shows the $\beta$ spectrum. One recognizes the subdivision into three
distinct regions: the small-$\beta$ region with $q\bar{q}g$
production, the medium-$\beta$ region with transversely-produced 
$q\bar{q}$ pairs, and the large-$\beta$ region 
where the longitudinal part dominates.  The sum of all 
three contributions leads to a rather flat $\beta$ spectrum. 

\begin{center}
\epsfig{file=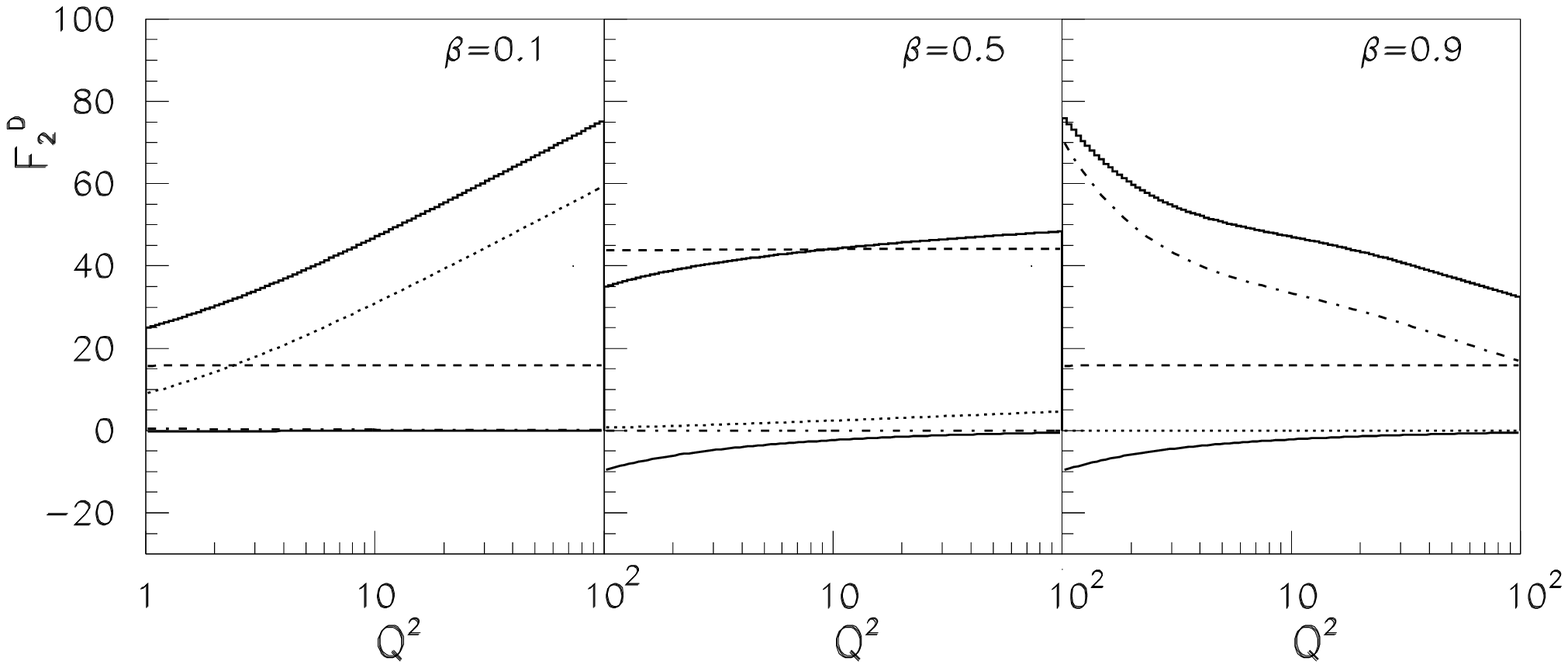,width=13cm}\\
Figure 6: The $Q^2$ distribution at fixed $x_{\fP}=0.001$. 
The notations of the lines are the same as in the previous figures. 
\vspace{0.3cm}\\
\end{center}

The $Q^2$ distribution in Fig.6 has the expected shape:
it is logarithmically increasing at $\beta=0.1$ (which we associate with 
$q\bar{q}g$ production),
constant for the leading-twist transverse part
at $\beta=0.5$, and decreasing for the longitudinal part
at $\beta=0.9$. The negative slope of the longitudinal contribution,
which is due to its higher-twist nature, is partly compensated by the 
counteracting logarithm in $Q^2$. The higher-twist correction for the 
transverse part, which is also decreasing when the absolute value is 
considered,
leads to a small positive slope at $\beta=0.5$ in the combined result. 

The transverse higher-twist correction to 
$q\bar{q}$ production wants to be negative, as expected from
theoretical arguments ~\cite{BWht}. Its absolute magnitude is rather small,
as can be seen in Figs.5 and 6, and it represents a correction to the three 
leading
pieces. As mentioned before, we have also attempted to include higher-twist
corrections to the $q\bar{q}$ production: with the available statistics 
it is not possible to assert whether such a contribution is present or not.

\begin{center}
\epsfig{file=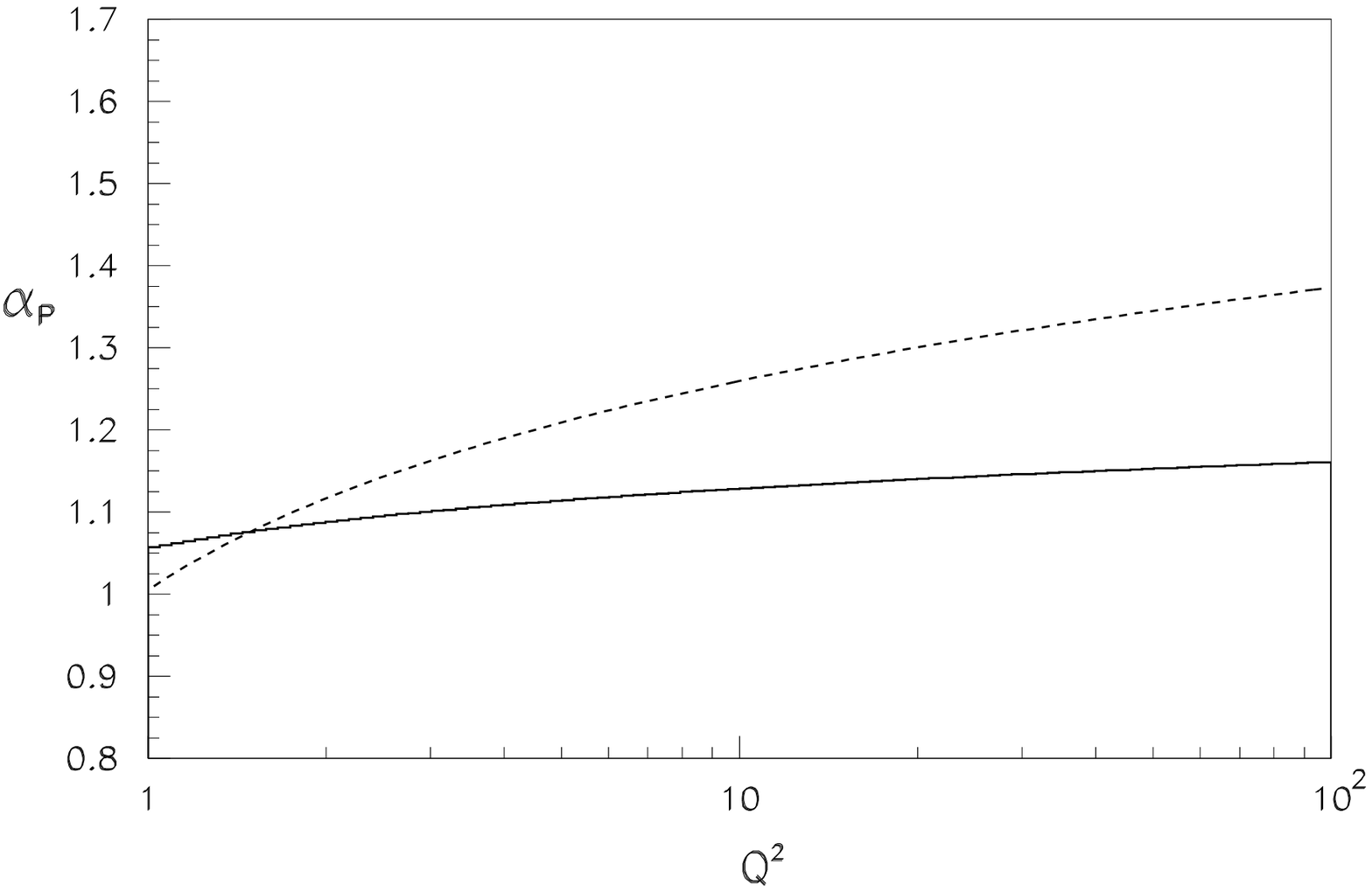,width=10cm}
\raisebox{3cm}{
\parbox{5cm}{Figure 7: Pomeron intercept for leading twist (solid line)
and higher twist (dashed line).}}
\vspace{0.3cm}\\
\end{center}

Although some of the parameters $n_{2\;0}$, ..., $n_{4\;1}$ have
substantial errors, their values, as determined from the fit, lead to an 
appealing theoretical scenario. 
The value $n_{2\;0}=1.11$ corresponds to an
effective Pomeron intercept $\alpha_{\fP}(0)=1.055$, which is consistent
with the `soft` Pomeron of Donnachie and Landshoff. It
exhibits, however, a slight rise with $Q^2$, which suggests that
in the leading-twist transverse cross section the Pomeron is already
a mixture of the soft Pomeron and the gluon structure function. According to
our discussion above, this means that the effective momentum scale of the 
aligned-jet configuration in $q\bar{q}$ production is slightly higher than
the 
typical hadronic scale of the soft Pomeron. 
The intercept for the higher-twist contribution (the dashed line in Fig. 7),
on the other hand, 
shows a strong rise with increasing $Q^2$. It tends towards a 
value of 1.2 at large $Q^2$, which is compatible with 
the gluon structure function, i.e., the Pomeron is hard in this case.

It is also interesting to compare the ZEUS fit with data from H1. Fig.8
shows all H1 data points with the ZEUS fit result overlaid.
The agreement is quite good when all points above $x_{\fP}=0.01$ 
are ignored. The strongest deviation is observed in the $\beta=0.2$ bin
at low $Q^2$. Here the ZEUS fit lies considerably above the H1 data. 
The overall impression, however, is that in the kinematic range 
where the Pomeron dominates over secondary trajectories, both data sets 
are consistent.
We find remarkable the good agreement between the H1 data and the ZEUS fit 
at $\beta=0.9$ and low $Q^2$, since
this region is not covered by ZEUS data. The fit seems to 
provide here a good extrapolation.
\begin{figure}
\begin{center}
\epsfig{file=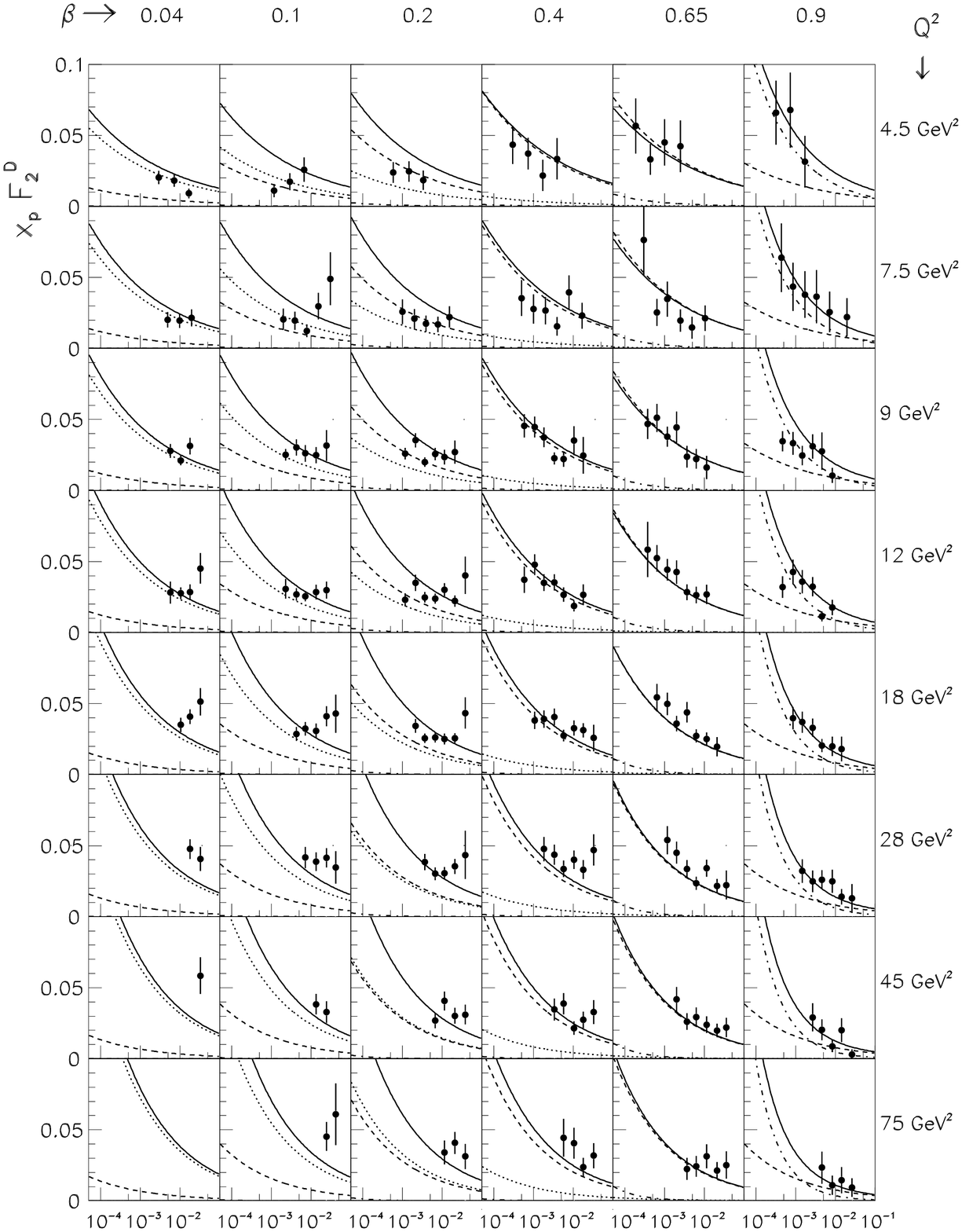,width=14cm}\\
\end{center}
Figure 8: The ZEUS fit compared to H1 data. Upper solid line: total result, 
dashed line: $F_{q\bar{q}}^T$,
dotted line: $F_{q\bar{q}g}^T$ and dashed-dotted line: $F_{q\bar{q}}^L$.
\end{figure}
In summary, we have shown that it is possible to describe the ZEUS data
using the first three terms in (12),the fourth term providing a small 
correction. In particular, there is no need for a hard gluon inside the
Pomeron. Furthermore, our fit is quite consistent with our 
theoretical expectations: the parameter $\gamma$ as well as the exponents 
$n_2$ and $n_4$ have chosen values which seem to confirm the ideas outlined 
in the previous section.

{\bf H1 Data:} Next we turn to the H1 data and describe the results of our
fit to them. Most remarkable is the fact that we find two different solutions.
The parameters of the first solution are given in Table 2, and the 
corresponding comparison of fit and data can be found in Fig.9.
\begin{table}[h]
\begin{center}
\begin{tabular}{|c|c|c|c|c|}   \hline
  & mean value & error & lower limit & upper limit \\ \hline
A & 16.8 & 8.4 & no & no\\ \hline
B & 13.9 & 4.2 & no & no \\ \hline
C & 10.5 & 2.6 & no & no \\ \hline
D &  0 & 6.5 & no & no \\ \hline
$x_0$ & 0.0033& 0.00063& 0.0001 & 1 \\ \hline
$\gamma$ & 0.28 & 0.08 & no & no \\ \hline
$n_{20}$ & 1 & 0.14& 1 & 2 \\ \hline
$n_{21}$ & 0.19 & 0.024& 0 & 1 \\ \hline
$n_{40}$ & 1.6 & 0.17 & 1 & 2 \\ \hline
$n_{41}$ & 0 & 0.82 & 0 & 1 \\ \hline
\hline$\chi^2$/d.o.f.   & \multicolumn{4}{c|}{139/130}\\  \hline
\end{tabular}\vspace{0.5cm}\\
Table 2
\end{center}
\end{table}
\begin{figure}
\begin{center}
\epsfig{file=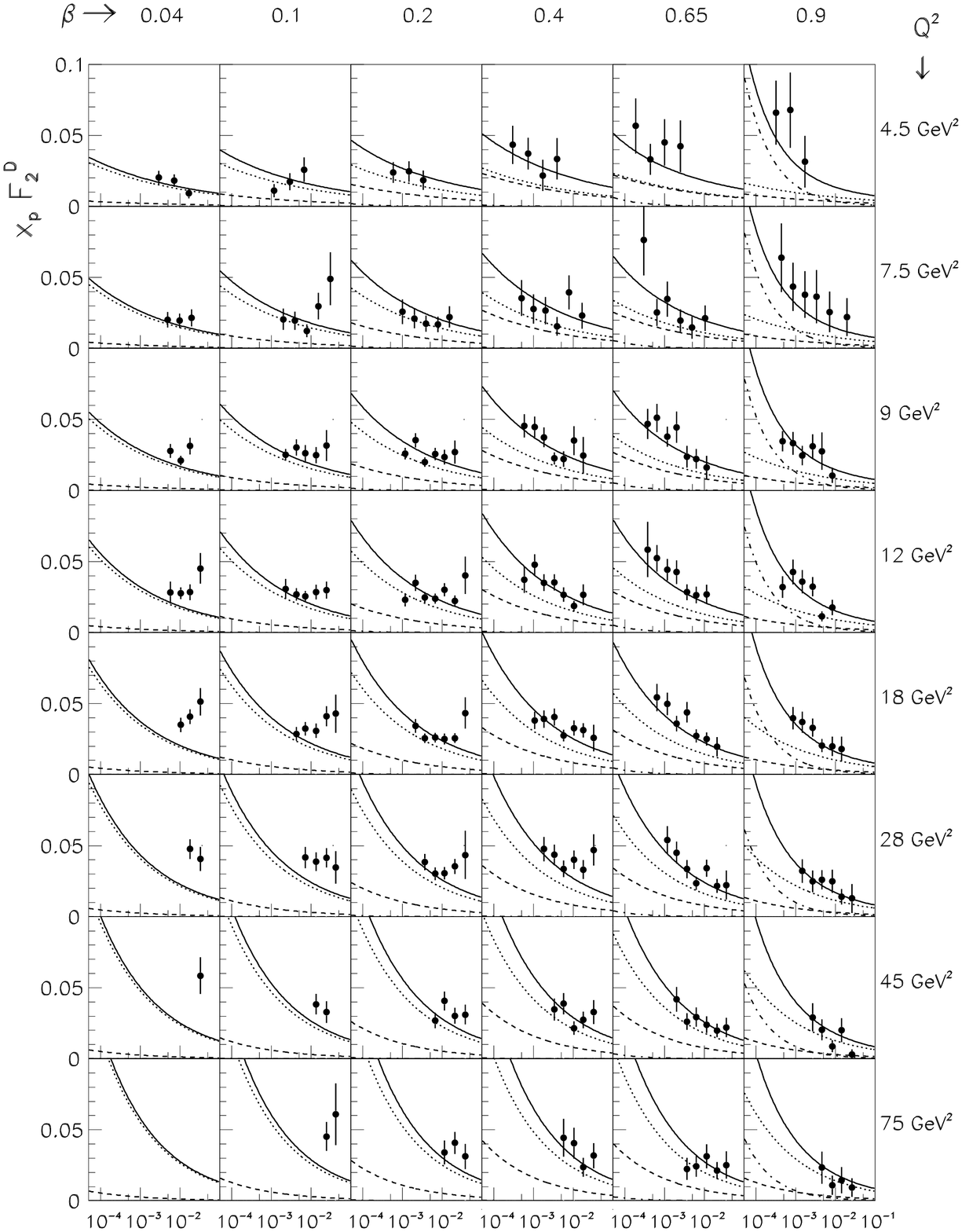,width=14cm}\\
\end{center}
Figure 9: The H1 fit compared to H1 data. Upper solid line: total result, 
dashed line: $F_{q\bar{q}}^T$,
dotted line: $F_{q\bar{q}g}^T$ and dashed-dotted line: $F_{q\bar{q}}^L$.
\end{figure}
In order to avoid secondary trajectories, 
we have imposed an upper cut on $x_{\fP}$
of 0.01. A striking feature of this solution is the fact 
that the exponent $\gamma$ is well below 1, which
can be interpreted as implying an initial gluon distribution 
that is singular for $\beta=1$. The explanation for this result 
can be inferred from Fig.10, where we have compiled all relevant diagrams, 
i.e., the $\beta$ spectrum, the $Q^2$ dependence and $\alpha_{\fP}$, in one
figure. 
\begin{center}
\mbox{
\raisebox{1cm}{\epsfig{file=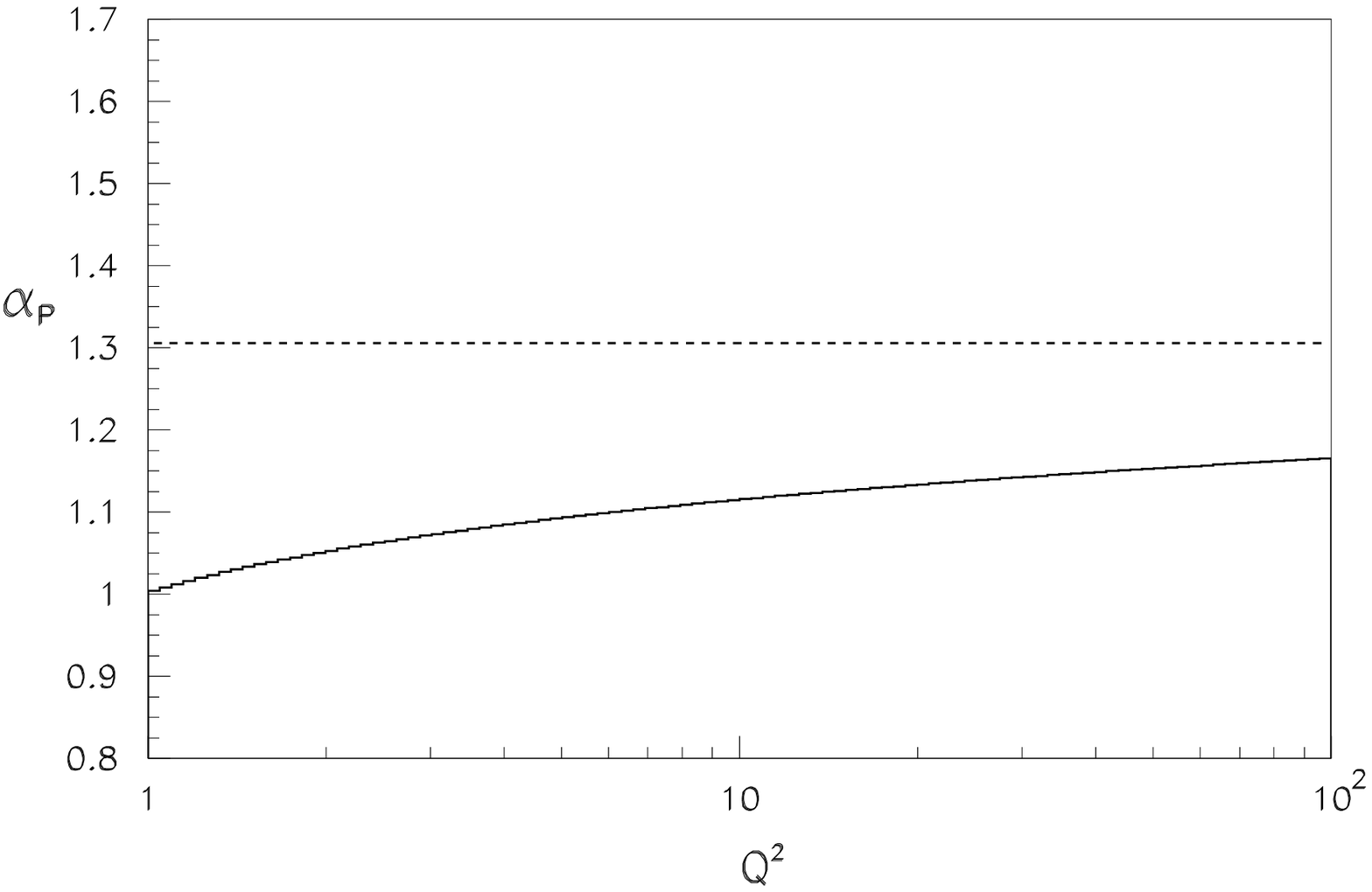,width=8.5cm}}
\epsfig{file=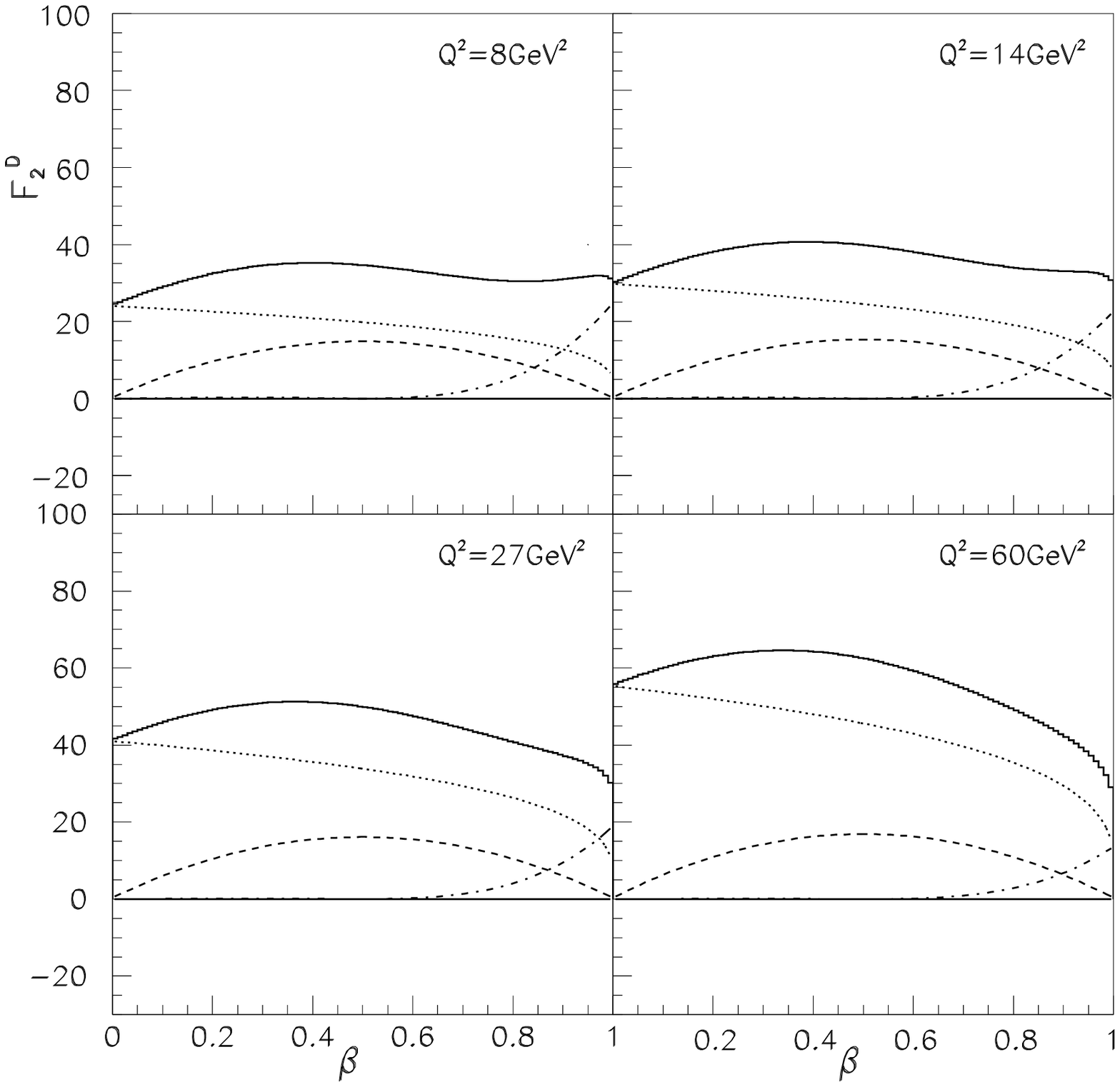,width=8.5cm}}
\epsfig{file=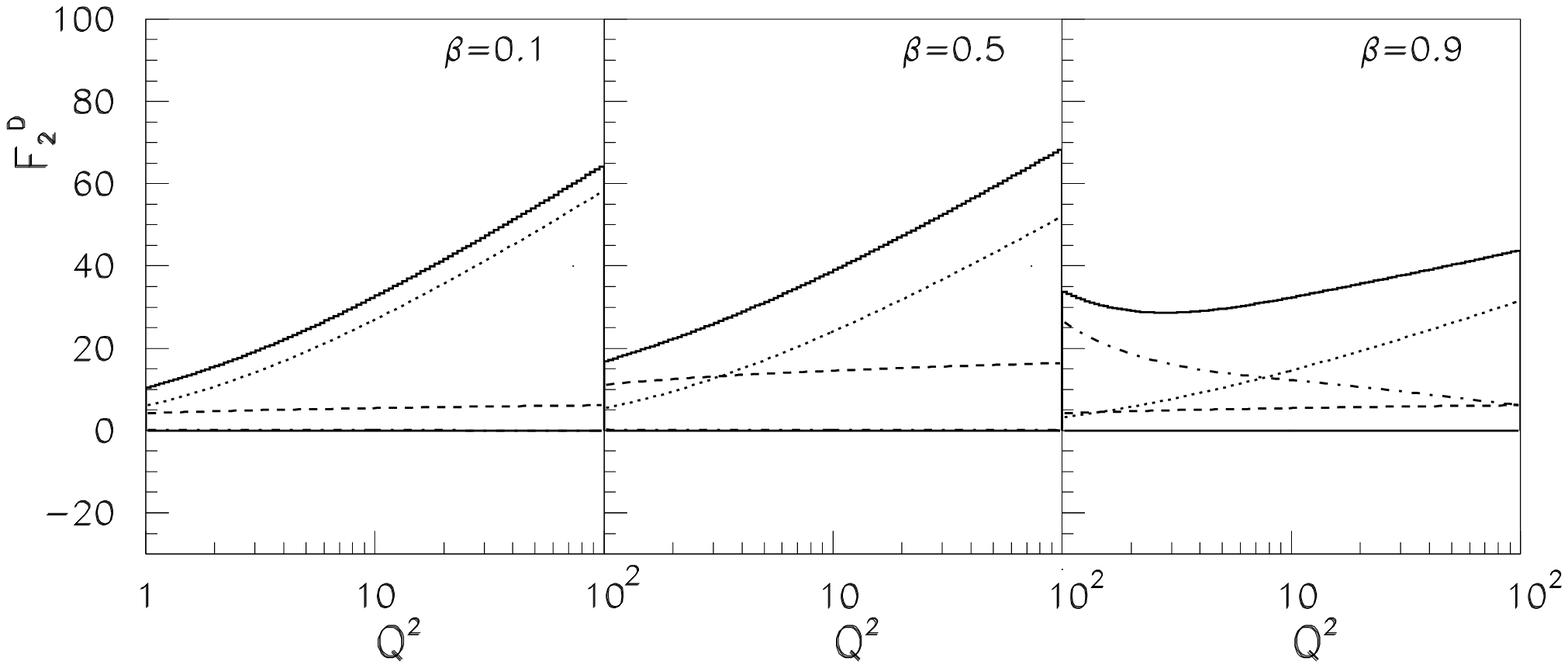,width=11cm}\\
Figure 10: The H1 fit results, in the same notation as in previous figures.
\vspace{0.3cm}\\
\end{center}
The H1 data prefer a positive slope in $Q^2$, even at a large $\beta$, of
0.5. With the present parametrization, this can only be achieved by making
the gluon contribution ($F^T_{q\bar{q}g}$) large. Of course, our 
rather simple approach is not competitive with the more sophisticated
analysis that has been performed by H1~\cite{H1}. Nevertheless, it
mimics the
\begin{figure}
\begin{center}
\epsfig{file=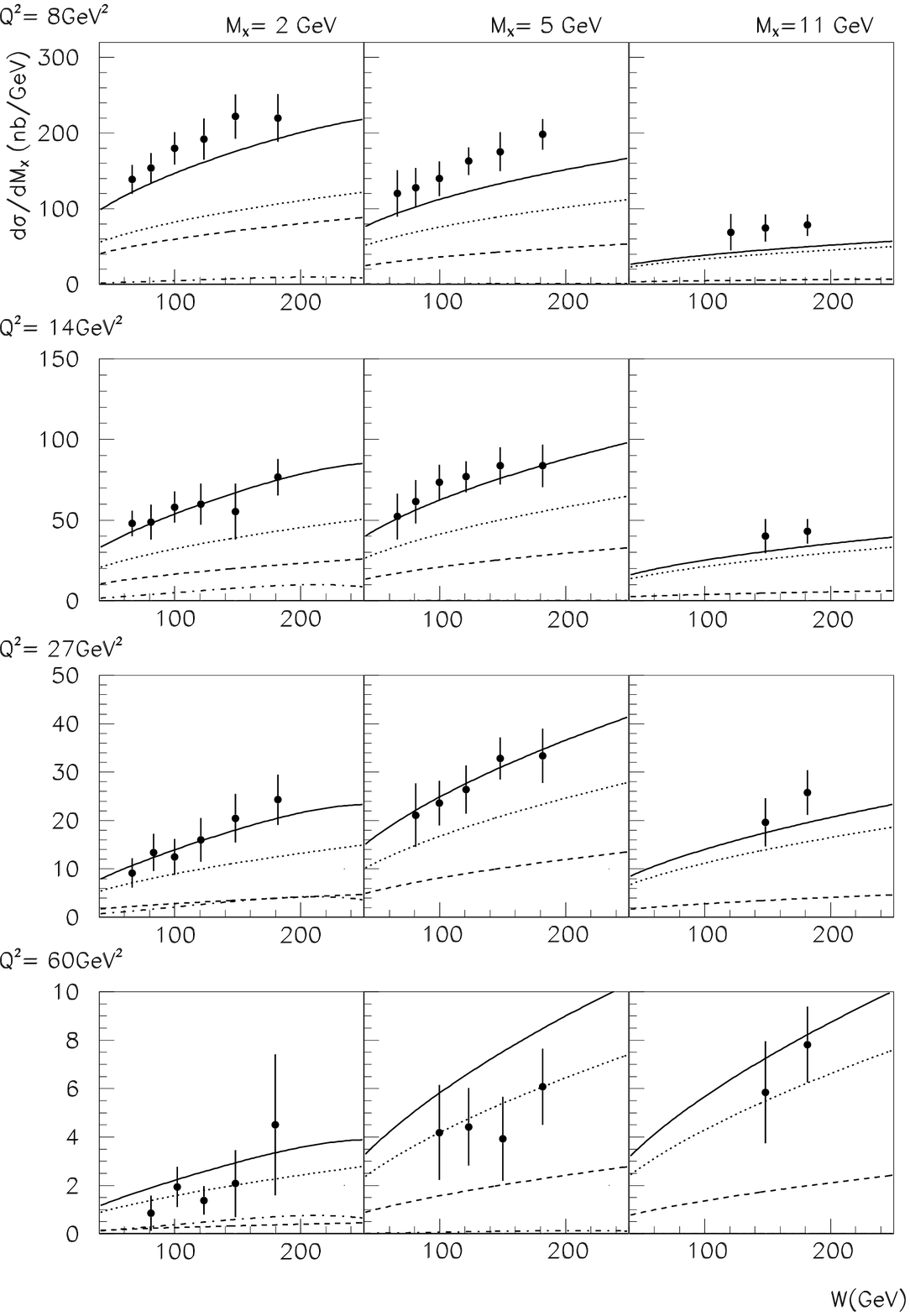,width=14cm}\\
\end{center}
Figure 11: The H1 fit compared to ZEUS data. Upper solid line: total result, 
dashed line: $F_{q\bar{q}}^T$,
dotted line: $F_{q\bar{q}g}^T$ and dashed-dotted line: $F_{q\bar{q}}^L$.
\end{figure}
effect of an evolving singular gluon distribution rather well. 
The $\beta$ spectrum in Fig.10 shows the dominance of the 
$q\bar{q}g$ contribution over the transverse $q\bar{q}$ contribution, which 
also spreads into the large-$\beta$ region. The longitudinal 
contribution has not disappeared completely, but is roughly a 
factor 2 smaller than in the fit to the ZEUS data. Since the  
$q\bar{q}g$ contribution has only a rather low Pomeron intercept associated
with leading twist, the intercept for the higher-twist contribution
is forced to a very high value of 1.3, in order to accommodate the 
data. It is completely flat, because we demanded a positive slope in our fit.
Without a lower limit on $n_{41}$, the slope would become negative.
 
Comparing with the ZEUS data, we find again reasonable agreement between
the two data sets. The most significant deviation is found for the lower 
$Q^2$ bins. We point out that, when one fits the ZEUS data
starting from the H1 data as input, the parameter values 
always move back towards those found in the earlier fit,
i.e., those in Table 1. In particular, the ZEUS data
do not favour a value of $\gamma$ smaller than $1$. If one requires a small
value of $\gamma$, one obtains an acceptable fit only with an unreasonably 
small value for $x_0$. If, in addition, $x_0$ is fixed at, say, $10^{-3}$, 
the $\chi^2$ value of the fit is substantially increased. 

As we have already said, the H1 fit has a second local minimum for a large 
$\gamma$ of 8.5. The $\chi^2$ is 
not much worse than in the previous fit (150 compared to 139).
The parameters if this solution are given in Table 3, and the comparison
with data is shown in Figs.12 and 13. The $\beta$ and $Q^2$ distributions 
are shown in Fig.14. On the whole, the distributions look similar
to the ZEUS fit. In more detail, however, there are a few differences: 
the $Q^2$ shape for $\beta=0.5$ is completely flat, and 
the $Q^2$ dependence of $n_4$ is also flat.

\begin{table}[h]
\begin{center}
\begin{tabular}{|c|c|c|c|c|}   \hline
  & mean value & error & lower limit & upper limit \\ \hline
A & 1865 & 451 & no & no\\ \hline
B & 1024 & 1.41 & no & no \\ \hline
C & 422 & 1.41 & no & no \\ \hline
D &  0 & 7.11 & 0 & no \\ \hline
$x_0$ & 0.0002& 0.000012& 0.0001 & 1 \\ \hline
$\gamma$ & 8.55 & 0.8 & no & no \\ \hline
$n_{20}$ & 1.16 & 0.026& 1 & 2 \\ \hline
$n_{21}$ & 0 & 0.047& 0 & 1 \\ \hline
$n_{40}$ & 1.44 & 0.093 & 1 & 2 \\ \hline
$n_{41}$ & 0 & 0.11 & 0 & 1 \\ \hline
\hline$\chi^2$/d.o.f.   & \multicolumn{4}{c|}{150/130}\\  \hline
\end{tabular}\vspace{0.5cm}\\
Table 3
\end{center}
\end{table}
Summarizing the H1 fit, we find when fitting our model to the H1 data 
a solution which allows for 
the singular gluon interpretation, but there is also a second solution
that is close to our model, i.e., consistent with the interpretation 
described in the context of the 
ZEUS data. Since the $\chi^2$ values of both solutions are not that 
different from each other, one has to search for further consistency checks:  
more decisive tests might be provided by comparisons with the vector-meson 
production cross section or with hard diffractive jets.
A singular gluon would give a strong transverse component, whereas in our 
model the longitudinal component is more pronounced. 
\begin{figure}
\begin{center}
\epsfig{file=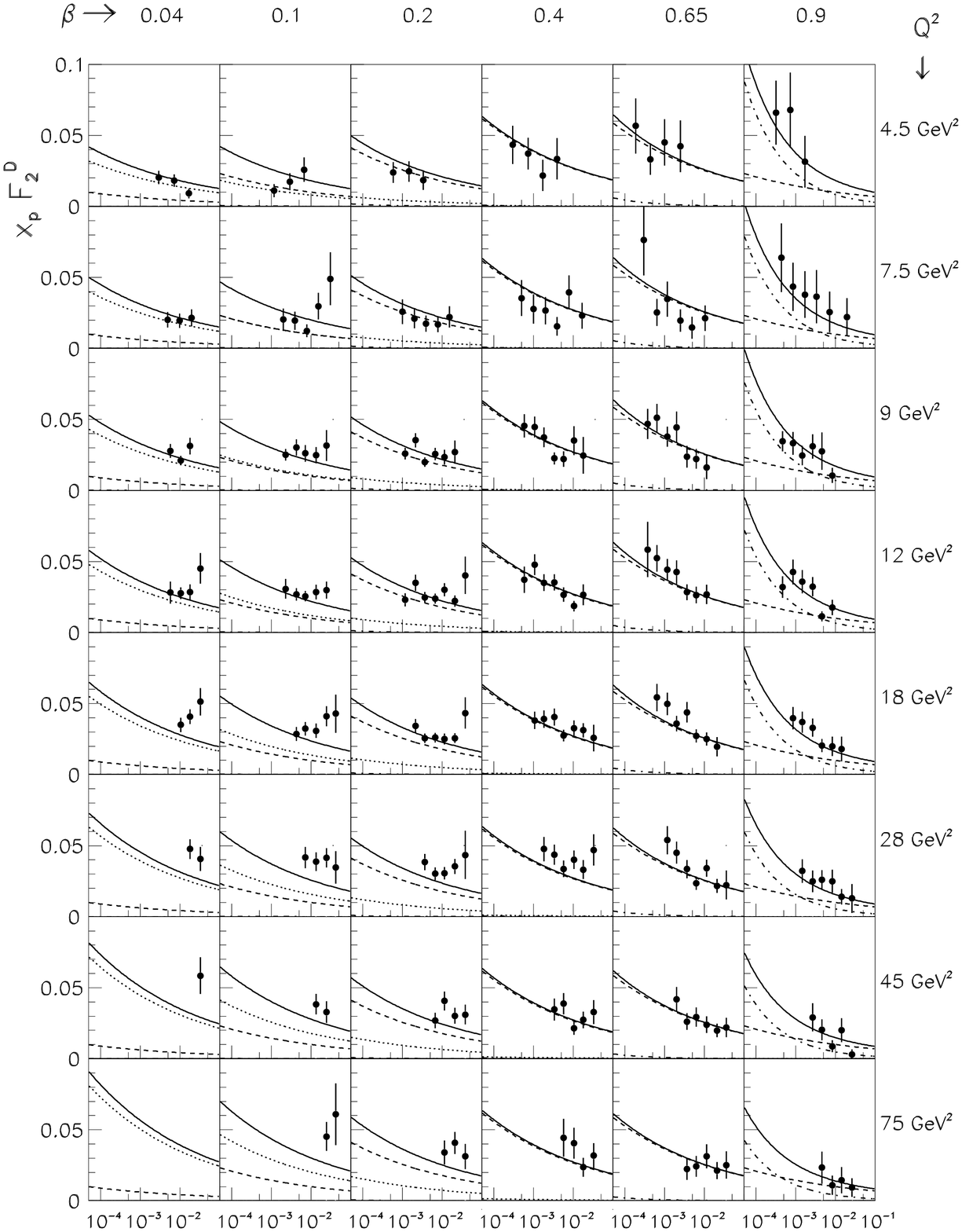,width=14cm}\\
Figure 12: The second H1 fit, with the same notations as before.
\end{center}
\end{figure}
\begin{figure}
\begin{center}
\epsfig{file=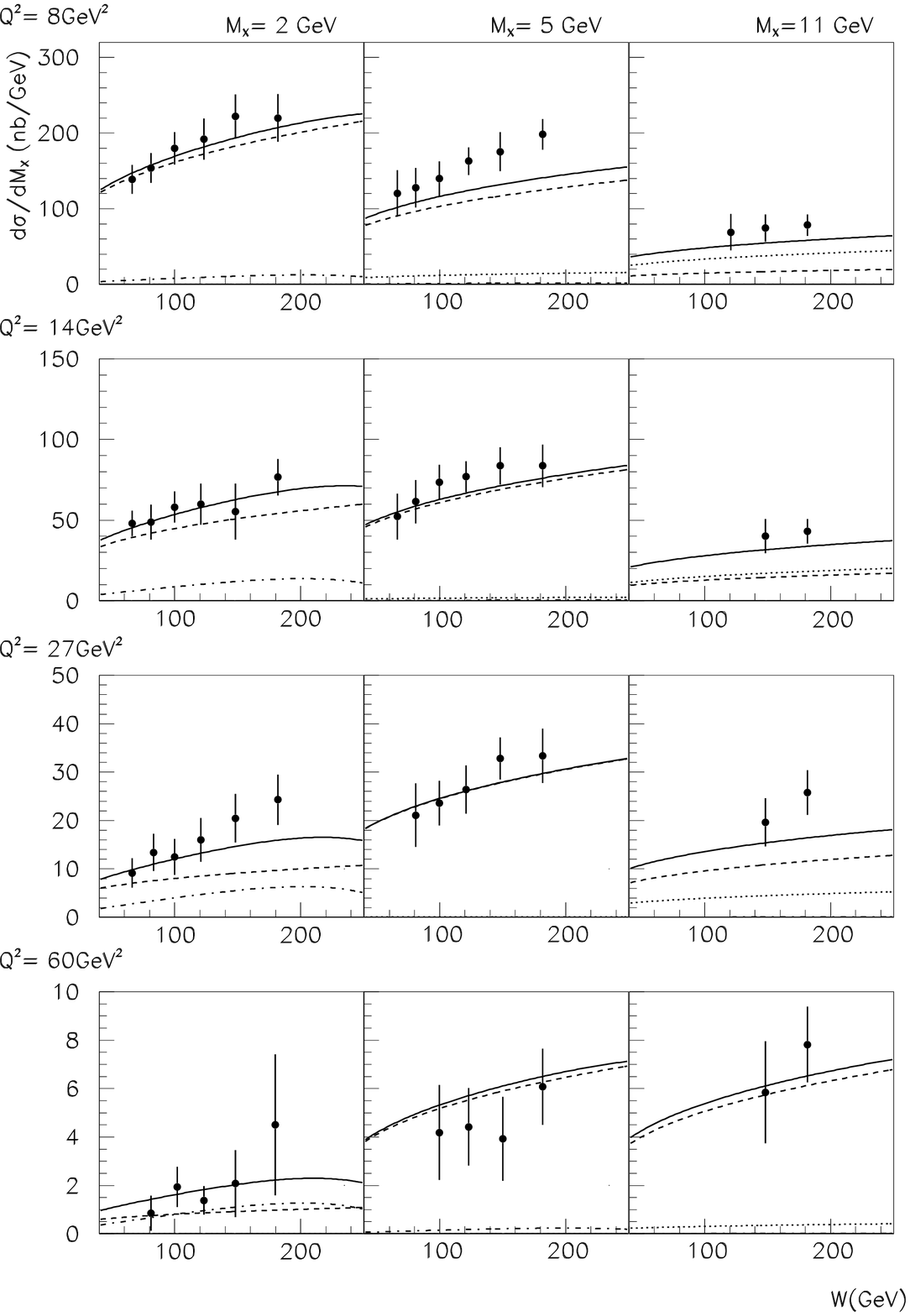,width=14cm}\\
Figure 13: The second H1 fit compared to the ZEUS data.
\end{center}
\end{figure}
\begin{center}
\mbox{
\raisebox{1cm}{\epsfig{file=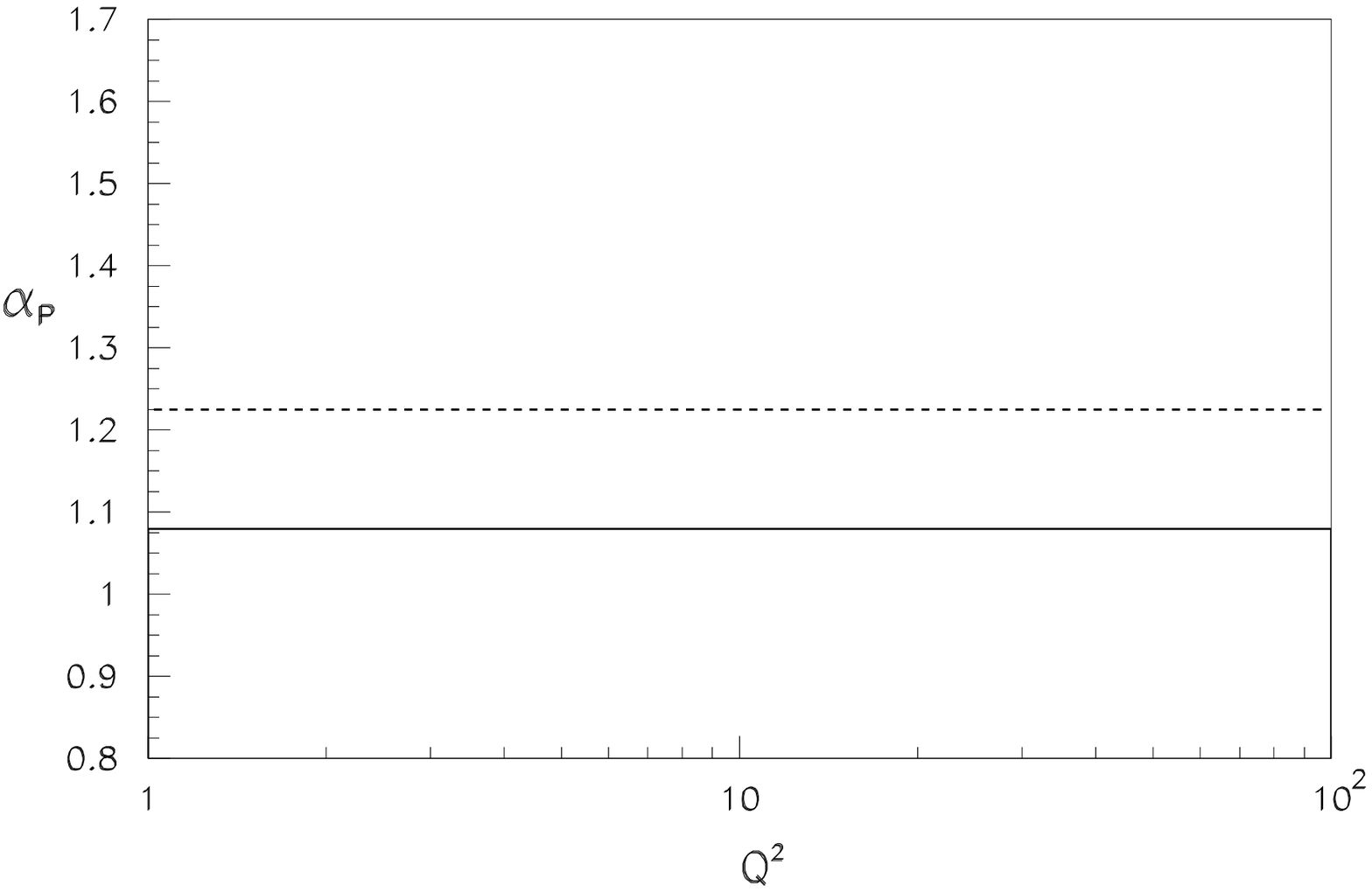,width=8.5cm}}
\epsfig{file=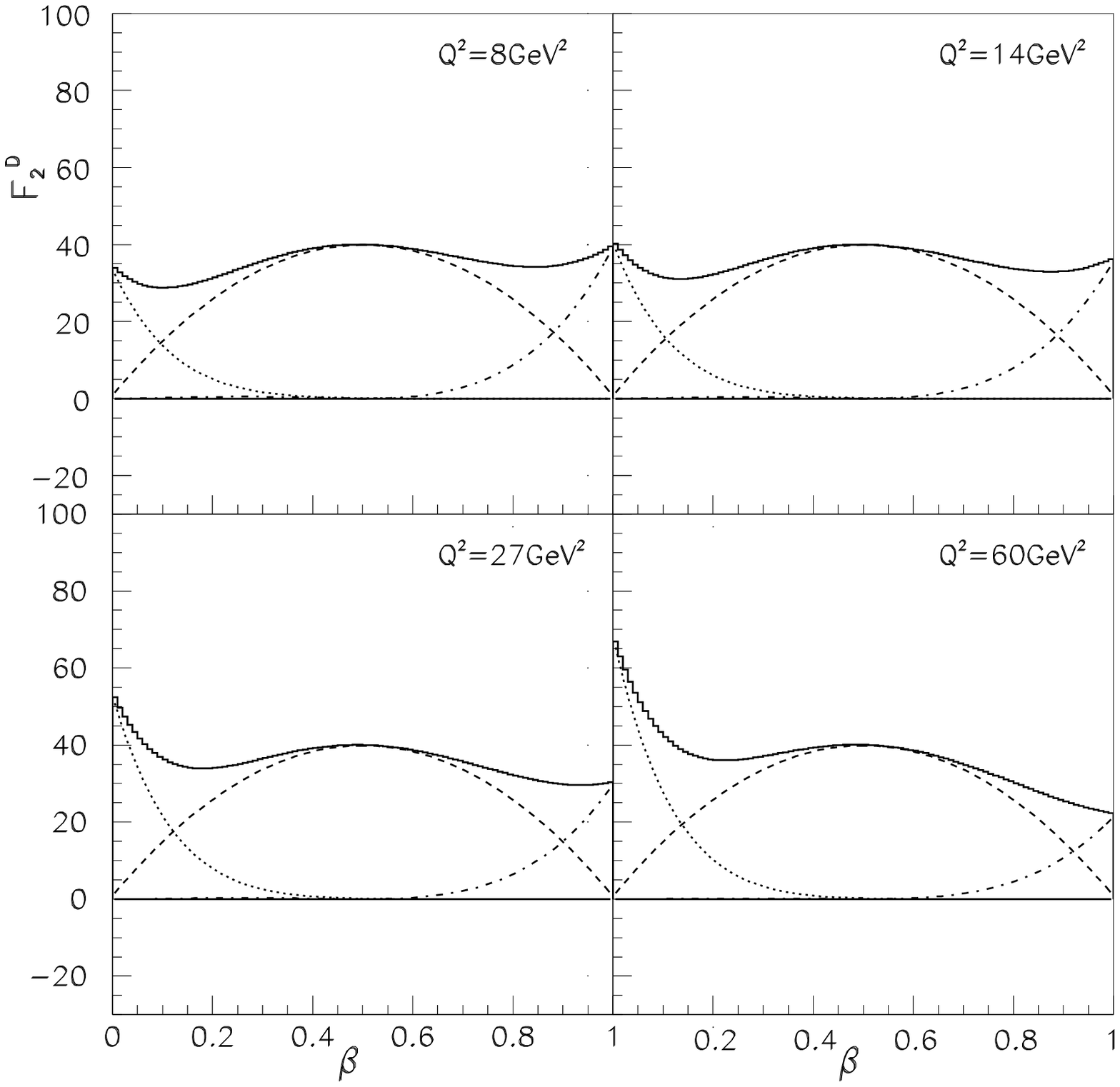,width=8.5cm}}
\epsfig{file=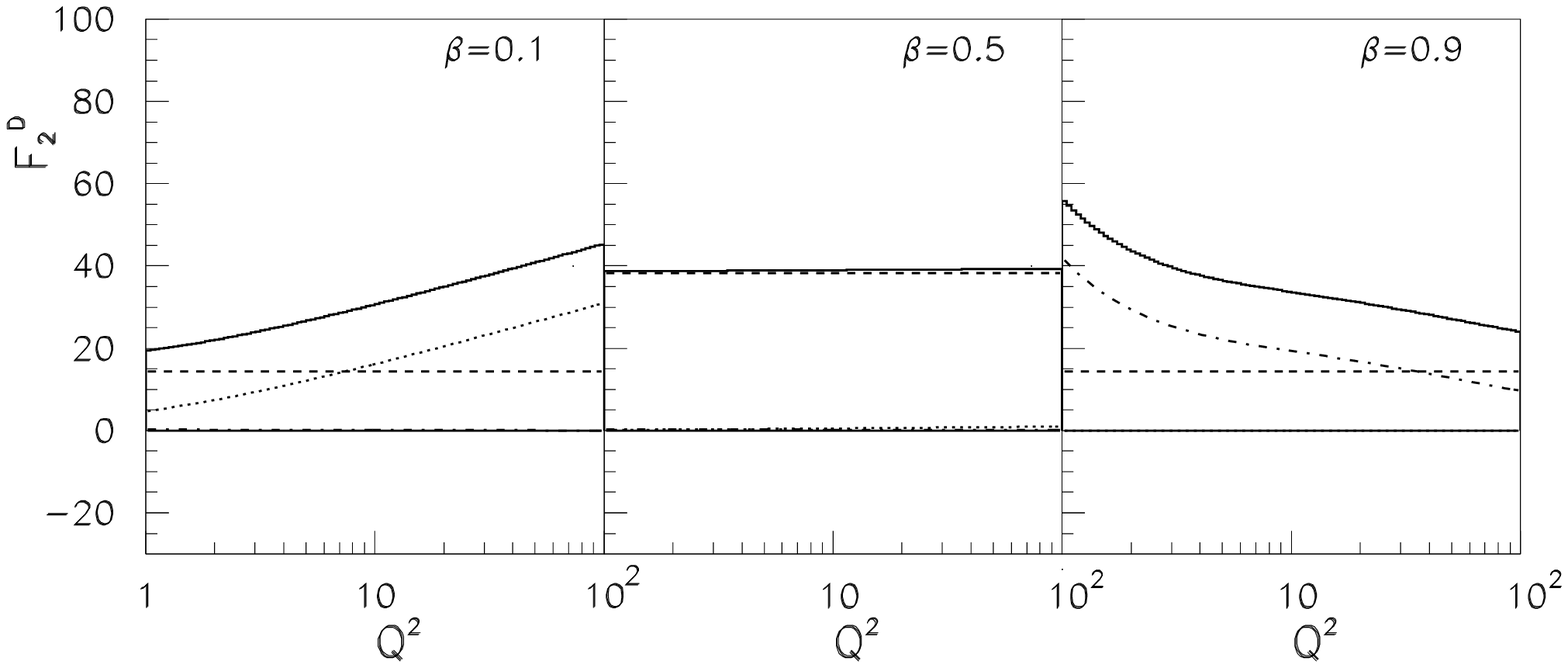,width=11cm}\\
Figure 14: The results of the second H1 fit.
\vspace{0.3cm}\\
\end{center}

\section{Conclusions}
In this analysis of diffraction in deep-inelastic scattering at HERA,
we have proposed a simple parametrization. Working in the wave-function
formalism and starting from 
the `hard part' of the diffractive cross section, we 
have suggested a simple
extrapolation into the `soft' nonperturbative region.
The main feature of the parameterization
is the decomposition of the $\beta$ spectrum into three contributions
which reside in separate regions with only
little overlap: $q\bar{q}g$ production at low $\beta$, 
transverse $q\bar{q}$ production at medium $\beta$, and longitudinal
$q\bar{q}$ production at large $\beta$. 
These can be derived from the corresponding wave functions
for the $q\bar{q}g$ and $q\bar{q}$ Fock states. The longitudinal contribution
is higher twist, whereas the other two contributions are leading twist.
The very different $\beta$ dependence of these three contributions 
allows to determine their relative importance through the fits to the data.
We fitted therefore our model to the recent ZEUS and H1 data.

An important result of the fit to the ZEUS data is the dominance of the
Pomeron-quark coupling at larger $\beta$ values ($\beta > 0.3$). The 
Pomeron-gluon coupling becomes substantial at lower $\beta$ ($\beta < 0.3$). 
The region of large $\beta$ ($\beta > 0.9$) is dominated by the longitudinal
contribution. 

The fit to the H1 data leads to two solutions. In order to be open towards
the H1 conjecture of a singular gluon distribution at
$\beta=1$~\cite{H1},
we have allowed, in the second term of our parametrization, the exponent 
$\gamma$ 
to be variable. A singular gluon distribution
would predict $\gamma<1$, whereas our model suggests $\gamma=3$,
or even a little bit higher when $Q^2$ evolution is included. Whereas the 
ZEUS fit found
only the option of a large $\gamma$ value, the 
two H1 solutions have very different $\gamma$
values.
In the first fit, $\gamma$ takes a small value, consistent with
the H1 conjecture. The Pomeron 
couples predominantly to gluons and only weakly to quarks. In the second 
solution, which has almost the same probability ($\chi^2$ value) as the 
first one, $\gamma$ is much larger
and the Pomeron couples mostly to quarks. The coupling to gluons becomes
substantial only at low $\beta$, quite similar to the ZEUS fit.

So far one cannot draw a final conclusion about the 
correct interpretation of
both data sets. On the one hand, the conjecture of a singular gluon
seems unlikely, in view of the ZEUS data. On the other hand, our 
parametrization with the two-gluon exchange as a model for the Pomeron 
can describe both data sets.
More insight into the final state is needed, which could, for example, 
be provided by careful analyses of vector meson 
production and/or diffractively-produced jets.

\end{document}